**Title:** Competition Between Controllable Non-Radiative and Intrinsic Radiative Second-Order Recombination in Halide Perovskites

**Authors:** Dengyang Guo[1,2], Alan R. Bowman[1,2,3], Sebastian Gorgon[1], Changsoon Cho[1,4,5], Youngkwang Jung[2], Jiashang Zhao[6], Linjie Dai[1,2], Jaewang Park[7], Kyung Mun Yeom[8], Satyawan Nagane[2], Stuart Macpherson[1], Weidong Xu[2], Jun Hong Noh[8,9], Sang Il Seok[7], Tom Savenije[6], Samuel D. Stranks[1,2]*

**Affiliations:**
1 Department of Physics, Cavendish Laboratory, University of Cambridge, Cambridge, UK.
2 Department of Chemical Engineering and Biotechnology, University of Cambridge, Cambridge, UK.
3 Laboratory of Nanoscience for Energy Technologies, STI, École Polytechnique Fédérale de Lausanne, Lausanne, Switzerland.
4 Department of Material Science and Engineering, Pohang University of Science and Technology (POSTECH), Pohang, Republic of Korea
5 Institute for Convergence Research and Education in Advanced Technology, Yonsei University, Seoul, Republic of Korea
6 Department of Chemical Engineering, Delft University of Technology, Delft, the Netherlands.
7 Department of Energy and Chemical Engineering, Ulsan National Institute of Science and Technology (UNIST), Ulsan, Korea
8 School of Civil, Environmental and Architectural Engineering, Korea University, Seoul, Republic of Korea.
9 Department of Integrative Energy Engineering, Korea University & Graduate School of Energy and Environment (KU-KIST Green School), Korea University, Seoul, Republic of Korea.

**Abstract**

Halide perovskite solar cells have demonstrated a rapid increase in power conversion efficiencies. Understanding and mitigating remaining carrier losses in halide perovskites is now crucial to enable further increases to approach their practical efficiency limits. Whilst the most widely known non-radiative recombination from solar cells relates to carrier trapping and is first order in carrier density, recent reports have revealed a non-radiative pathway that is second order. However, the origin and impact of this second-order process on devices remain unclear. Here, we understand this non-radiative second-order recombination ($k_{2non}$) pathway by manipulating the charge carrier dynamics via controlling the bulk and surface conditions. By combining temperature-dependent spectroscopies, we demonstrate that the value of $k_{2non}$ depends on extrinsic factors, in contrast to intrinsic second-order recombination, which aligns with theoretical evaluations through van Roosbroeck-Shockley relations. Based on density functional theory simulations and Quasi-Fermi level calculations, we propose that shallow surface states are the primary origin of this second-order non-radiative component, contributing up to ~80 mV of the overall reduction in $V_{oc}$ at room temperature. This work reveals that carrier losses from two non-radiative recombination types (first and second order) are not linked, emphasizing the need for distinctive mitigation strategies targeting each type to unlock the full efficiency potential of perovskite solar cells.

I. INTRODUCTION

Perovskite solar cells have emerged as a promising candidate for next-generation photovoltaics due to their excellent optoelectronic properties and rapid progress in power conversion efficiencies. With reported efficiencies now reaching 27%,[1] perovskite solar cells have already demonstrated their potential to compete with efficiencies of traditional silicon-based technologies. To drive down power conversion losses to their lowest possible values and approach their practical limits, a complete understanding of charge carrier recombination dynamics for solar cells is crucial, regardless of the choice of absorber materials.

In all solar cells, the theoretical maximum efficiency—governed by the Shockley-Queisser (SQ) limit—assumes idealized conditions, including perfect charge extraction and no non-radiative recombination losses. [2–5] However, in practical devices, material imperfections in the absorber layer (e.g., bulk and surface recombination) impose a lower, practical efficiency limit. This limit reflects the best performance achievable with existing absorber materials under realistic constraints. While optimizing charge transport layers and electrode contacts can reduce resistive losses and improve device performance, such enhancements cannot surpass the practical efficiency limit inherently defined by the absorber's non-ideal properties. Thus, advancing the absorber material itself remains critical for bridging the gap between practical and theoretical efficiencies.[2,6] These losses are attributed to non-radiative carrier recombination, which can typically (for example in halide perovskites) be categorized as first-order, second-order, or third-order processes, depending on the excitation density. Typically, non-radiative carrier losses in halide perovskites have been linked to first-order trap-assisted recombination, which is dominant at low charge carrier densities and characterized by coefficient $k_1$, or to third-order Auger recombination, which prevails at high charge carrier densities (typically beyond the operational regimes of solar cells). However, at operating charge carrier densities, second-order processes are frequently dominant (especially once first order charge trapping has been minimised) and typically considered radiative, represented by coefficient $k_{2r}$. This process is often referred to as band-to-band or radiative recombination between free electrons and holes, an intrinsic property of semiconductors, while first-order processes depend on extrinsic factors.

However, recent investigations by us and others have revealed the presence of non-radiative second-order recombination processes[7–10], which we herein define with coefficient $k_{2non}$. Such a presence means the overall second-order recombination with rate $k_2$ is not equal to the radiative rate $k_{2r}$, but rather $k_2 = \eta_{esc} k_{2r} + k_{2non}$, accounting for both radiative and non-radiative second-order processes. Here, $\eta_{esc}$ stands for the escape probability of photons from the film, including the effect from photon recycling. However, the second-order non-radiative recombination processes remain unexplained, and it is unclear whether they relate to intrinsic

(and thus unavoidable) or extrinsic (and potentially controllable) processes. Nonetheless, they are regarded as the reason for a drop of ~50-100 mV in the open-circuit voltage ($V_{oc}$) of perovskite solar cells.[11] Therefore, understanding this recombination channel is crucial to guide strategies aimed at achieving the highest performance levels. For instance, elucidating the interplay between these non-typical second-order losses and classical first-order recombination will help in targeting the minimization of both losses either independently or simultaneously to enhance perovskite device performance.

Here, we present an approach to yield precise recombination parameters and, together with temperature-dependent measurements, we extract both $k_1$ and $k_{2non}$ for various high-performance formamidinium lead iodide (FAPbI$_3$) variants. Using surface passivation methods as a lever, we demonstrate that the change in the radiative second-order recombination rate with temperature is in line with theoretical evaluated values. By contrast, we show that the absolute value of $k_{2non}$ is an extrinsic parameter that relates to the surface, and the temperature-dependence of the values varies between samples. These findings are supported by Density Functional Theory (DFT) calculations, which support the proposition that surface shallow states could contribute to $k_{2non}$. Finally, utilising the temperature-dependent recombination parameters, we calculate the practical impact on carrier losses for solar cell device open circuit voltage, $V_{oc}$, from sources of non-radiative recombinations via either $k_1$ or $k_{2non}$. The $V_{oc}$ drop due to $k_{2non}$ highlights the importance of eliminating the controllable $k_{2non}$, in addition to the more widely known and controllable $k_1$, when striving to reach the practical efficiency limits for perovskite solar cells.

## II. RESULTS AND DISCUSSION

### A. A simple but accurate fitting approach reveals k$_{2non}$

A useful method to check film quality is by measuring recombination rates via time-resolved spectroscopy, such as time-resolved photoluminescence (TRPL). In several reports, arbitrary multi-exponential fitting is applied to the data to obtain a combination of two rates, but such approaches lack physical meaning (explained in supplementary Note 1) and thus are incapable of providing real recombination rates. Here, we employed an approach that is physically reasonable for representing the recombination processes and is mathematically logical for extracting the parameters. The rate equation for a photo-excited carrier population $n$, assuming uniform carrier density, can be described using a simplified equation:

$$\frac{dn}{dt} = G - k_1 n - k_2 n^2 - k_3 n^3 \tag{1}$$

Here, the $k_1$, $k_2$, and $k_3$ represent the first, second and third order (Auger) recombination rates, respectively. $G$ is the generation term, which is time-independent for continuous excitation and time-dependent for pulsed laser excitation. Our measurements below explore two measurement conditions: pulsed excitation[12] and steady-state excitation[13]. Under transient excitation, we solve Equation 1 to obtain the PL intensity as a function of time[12],

$$PL(t) = c\left(\frac{n_{init}k_1}{n_{init}k_2(e^{k_1 t}-1)+k_1 e^{k_1 t}}\right)^2 \tag{2}$$

Here, we neglect the Auger recombination term, which is negligible[14] under the excitation conditions used in our experiments. The $n_{init}$ is the excited carrier density calculated by the recorded incident fluence, beam size and film thickness. The constant $c$ contains information about the photon collection efficiency of the instrument. We also assume lateral uniform excitation density over the sample and negligible charge transport effects for the time window over thousands of nanoseconds, supported by the ultrafast propagation and long diffusion lengths of carriers in perovskites[15–17].

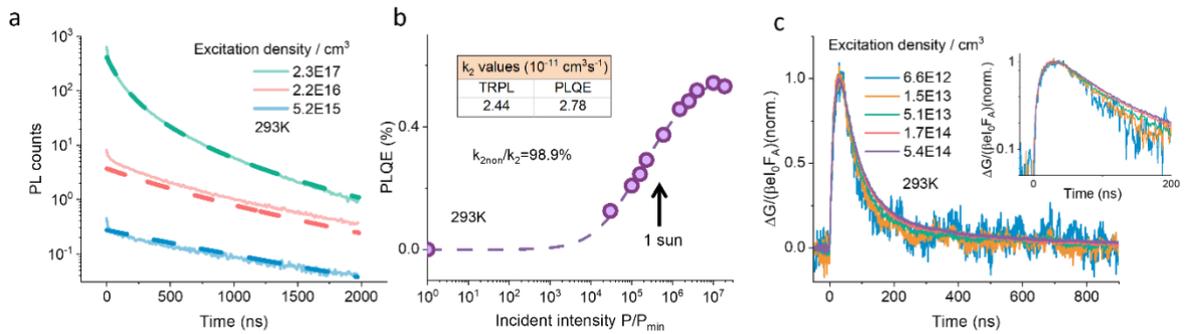

*FIG. 1. Fits of TRPL and PLQE data acquired on thin films of FAPbI$_3$ EDTA-24h at room temperature. The fits exhibit agreement in the value of the overall second-order recombination rate, $k_2$, despite the different measurements and fitting models. (a) Excitation-density-dependent TRPL (solid lines) and fits (dashed lines). The excitation wavelength is 520 nm, and the repetition rate is 1 kHz. The $k_2$ value is double-checked by an exponential fitting of the trace at the lowest excitation density, shown in Fig. S1. (b) Excitation-density-dependent PLQE (solid symbols) and fits (dashed line) of PLQE using Equation 3. The excitation wavelength is also 520 nm under continuous wave (CW) mode. The arrow points at the excitation density close to standard 1 sun illumination. (c) Excitation-density-dependent TRMC traces normalised to 1, with excitation at 500 nm with a repetition rate of 10 Hz. Inset: log-linear plot of the traces.*

To measure and analyse recombination parameters, especially $k_{2non}$, we first deposit stabilised FAPbI$_3$ onto glass substrates, following a recipe using EDTA additives[18], and the film sample is then treated with post-annealing in air for 24 hours, and hereafter termed EDTA-24h. Next, we measure the TRPL decay curves with initial excitation densities generated from 520-nm-wavelength pulsed light from $5.2 \times 10^{15}$ cm$^{-3}$ to $2.3 \times 10^{17}$ cm$^{-3}$. As shown in Fig. 1a, we use Equation 2 for global fitting on the excitation density-dependent TRPL traces. Here, global fitting means we input the excitation densities in a single step to fit all three traces simultaneously and the only two variables are $k_1$ and $k_2$ as the variables, while $c$ is a common constant representing the photon collection efficiency of the instrument. This method ensures a valid $k_2$ by globally matching the decays under high (~$10^{17}$ cm$^{-3}$), medium (~$10^{16}$ cm$^{-3}$), and low (~$10^{15}$ cm$^{-3}$) excitation densities, including 1 sun illumination (equivalently between ~$10^{16}$~$10^{17}$ cm$^{-3}$) and representing recombination which is dominated by second-order, mixed-second and first-order recombination regions, respectively. Moreover, the rate constant $k_1$ obtained via the global fitting ($5.2 \times 10^5$ s$^{-1}$) is validated by an independently exponential fitting on the trace under the lowest excitation density ($4.6 \times 10^5$ s$^{-1}$, supplementary Note 1, Table S1). The close agreement between these methods confirms the robustness of the global fitting, even as the system's complexity necessitates a multi-parameter analysis.

From the above fitting, we obtain the overall $k_2$ ($2.44 \pm 0.03 \times 10^{-11}$ cm$^3$s$^{-1}$), which is the coefficient representing all second-order recombination. To explore the components of $k_2$, we employ our recently developed photoluminescence quantum efficiency (PLQE)-based approach[13]. Typically, $k_2$ is interpreted as radiative from band-to-band recombination and thus is a constant value representing an intrinsic direct bandgap semiconductor property. Hence, $k_{2r}$ and $k_2$ are not distinguished in the common definition of PLQE within the recombination model as outlined:

$$\eta_{PLQE} = \frac{\eta_{esc} k_{2r}(p_0 n + n^2)}{k_1 n + k_2 n^2 + k_3 n^3} \tag{3}$$

Here, $p_0$ is the background hole concentration, which is negligible compared to carrier densities under 1 sun illumination for most lead-based perovskites including those explored here (can be equally replaced in case of n-doping by $n_0$).

We use Equation 3 to fit the excitation density-dependent PLQE values, using a stochastic strategy, which considers the intrinsic randomness and statistical fluctuations in the parameters that govern PLQE values[13]. This method, applied to the PLQE values across a range of steady-state excitation densities spanning two orders of magnitude, extracts individual values for $k_2$ and $\eta_{esc} k_{2r}$. From the fitting as shown in Fig. 1b and Table S2, we obtain distinct values for $k_2$ ($2.78 \times 10^{-11}$ cm$^3$s$^{-1}$) and $\eta_{esc} k_{2r}$ ($0.02 \times 10^{-11}$ cm$^3$s$^{-1}$). While a hypothetical scenario could posit equality between these parameters (e.g., $k_2 = \eta_{esc} k_{2r}$) our unconstrained fitting procedure inherently yields non-identical values. This result indicates that such equality is neither required nor consistent with the data, allowing us to rule out this scenario. These values are listed in Table S1. Thereby, we confirm again that $k_2$ and $\eta_{esc} k_{2r}$ are different, subject to $k_2 = \eta_{esc} k_{2r} + k_{2non}$, accounting for both radiative and non-radiative processes. This is consistent with previous works in which distinct $k_2$ values are obtained by the same fitting method on transient absorption (TA) and TRPL, revealing the existence of the $k_{2non}$ component.[7] Herein, the $k_{2non}$ values are obtained through the extracted $k_2$ and $\eta_{esc} k_{2r}$ values from Equation 3.

While fitting PLQE data using Equation 3, we need to input the value of $k_1$ to extract $\eta_{esc} k_{2r}$. To ensure accuracy, in addition to cross-verifying the $k_1$ value from the global fits via Equation 2, we carried out time-resolved microwave photoconductivity (TRMC) measurements. These TRMC results confirm the first-order recombination regime, as illustrated in Fig. 1c. Compared to recording radiative recombination of carriers by PL, TRMC records the overall recombination of free charge carriers, regardless of whether they recombine non-radiatively or radiatively. As shown in Fig. 1c inset, the TRMC decays show an overlapping and straight behaviour under excitation densities in the range below $5.4 \times 10^{14}$ cm$^{-3}$, where TRPL is no longer accessible from the instrument used in this study. In this range, the carrier recombination becomes non-radiative between trapped electron and background holes, but with a rate ($2 \times 10^6$ s$^{-1}$) larger than $k_1$ obtained from PL, indicating that PL is more influenced by surface traps since TRMC monitors residual mobile carriers in the bulk[19]. We herein continue to use TRPL for fitting but use TRMC as a cross-check.

Overall, our methodology employs two complementary pathways to determine $k_2$: the first uses Equation 2 to globally fit TRPL data, providing a direct estimate of $k_2$ (alongside $k_1$, which is independently validated via exponential fitting), while the second and primary approach uses PLQE recordings (Equation 3), combined with the pre-determined $k_1$ values, to extract $k_2$ and further decouple it into its components, including the key parameter of interest, $k_{2non}$. While the TRPL fits serve as a useful cross-check for $k_2$, the PLQE + $k_1$ approach is the primary driver for obtaining $k_2$ and $k_{2non}$. Encouragingly, we find that, despite the different measurements and fitting models, the overall second-order recombination rate, $k_2$, obtained from the fitting approaches above is in good agreement: $(2.44 \pm 0.03)$ vs. $(2.78 \pm 0.32) \times 10^{-11}$ cm$^3$s$^{-1}$ from TRPL and PLQE fitting respectively, as shown in Fig. 1b. Such values are consistent with previous results of $k_2$ from various time-resolved spectroscopies, such as THz spectroscopy[20], TRMC[8,19], TA[7], and PL[7,13]. These results, consistently falling within the range of $\sim 10^{-11}$ – $10^{-10}$ cm$^3$s$^{-1}$, also align with simulations.[21] This alignment, however, masks the potential existence of a non-radiative component ($k_{2non}$) within the total $k_2$, as the consistency implies completeness in the measured values. By combining global TRPL fitting (directly providing $k_2$ for cross-check and $k_2$) with stochastic PLQE fitting (decoupling $k_2$ into $\eta_{esc} k_{2r}$ and $k_{2non}$ using $k_1$ values from global TRPL fitting, validated through exponential fitting), we extract $k_{2non}$ =$2.75 \times 10^{-11}$ cm$^3$s$^{-1}$ out of the total $k_2$ of $2.78 \times 10^{-11}$ cm$^3$s$^{-1}$. This reveals that $k_{2non}$ dominates the total $k_2$ (98.9%), further consistent with the sample's low PLQE of around 0.4% at the excitation density equivalent to 1 sun.

### B. The temperature-dependent analyses reveal that $k_{2non}$ values are independent of $k_1$ and relate to an extrinsic property.

To elucidate the origin of the non-radiative component of $k_2$ (i.e., $k_{2non}$), we carry out temperature-dependent measurements. As a prerequisite, we first examine the PL spectra of each sample under different temperatures, as shown in Fig. S2. The shift of the PL peak aligns with the change in the bandgap at lower temperatures, and the full width at half maximum (FWHM) of the PL peak becomes narrower. Importantly, no additional peak or shoulder emerges in the PL spectra at lower temperatures over the range studied here (50 K – 293 K, Fig. S2), indicating that it is suitable for temperature-dependent analysis through either TRPL or PLQE.

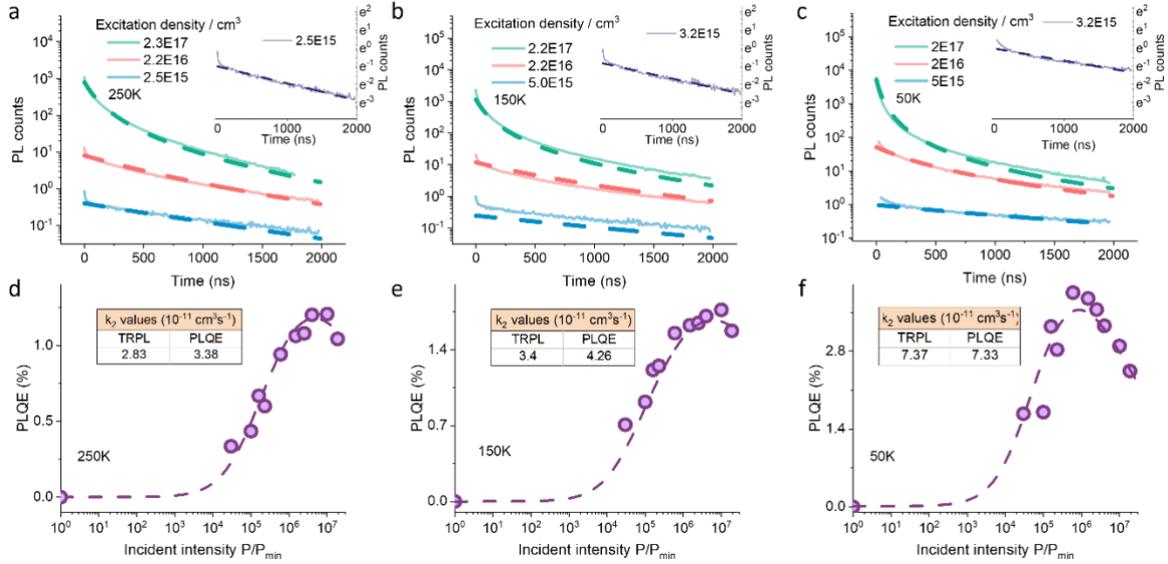

FIG. 2. Fits of TRPL and PLQE data obtained from FAPbI$_3$ EDTA-24h thin films at low temperatures. The estimated generation rate at the minimum excitation power, $P_{min}$ in (d-f) is around $7\times 10^{16}$ cm$^{-3}$s$^{-1}$. Consistent $k_2$ values are obtained from the fits (dashed lines) of TRPL (solid lines) and PLQE (solid symbols) at each temperature, following the method applied at room temperature in Fig. 1, using Equation 2 and 3, respectively. The temperature-dependent traces of TRPL (a-c) are from direct recordings while the low temperature PLQE (d-f) values are from scaling steady-state PL emission[13,22]. Results at other temperatures are shown in Fig. S3&4. Insets of (a-c): Exponential fits (dashed lines) to the TRPL recorded (solid lines) under the lowest excitation density in ln-linear plots, with the excitation density stated (cm$^{-3}$). These linear fits ensure the validity of the $k_1$ values obtained from the global fitting.

To extract the $k_2$ values from both the TRPL and PLQE approaches at each temperature, we first record and carry out global fitting of TRPL traces at low temperatures, using Equation 2, as shown in Fig. 2a-c. Then, we record the PL emission spectra under the same excitation and signal collection conditions. This ensures an accurate ratio of PL spectra integration between low temperature and room temperature. Given no change in absorbance on temperature at the excitation wavelength (Fig. S2), and assuming that the ratio between PL emissions at two temperatures represents the ratio of PLQE at those temperatures,[13] we record the PL spectra at the excitation densities equivalent to those used in PLQE records. This allows us to obtain the excitation-density-dependent PLQE at each temperature, as shown in Fig. 2 and Fig. S4. We adopt this scaling method because direct PLQE measurements at low temperatures using integrating spheres are in general not feasible. We observe across the different samples that the PL emission intensity and consequently the PLQE increases at lower temperature under the same excitation densities. Notably, most apparent at the lowest temperature, the excitation-density-dependent PLQE shows a downturn at high excitation densities, indicating the onset of Auger recombination. This onset is consistent with an increase in the Auger recombination rate and an increase in the carrier density due to reduced non-radiative recombination at low temperatures.[23]

At each temperature, we also record the TRMC traces to confirm first-order recombination behaviour under the lowest excitation density. The temperature-dependent TRMC decays are shown in Fig. S5 and mobilities are in Fig. S6. At low temperatures, the first-order recombination rate becomes smaller, which is also in line with the rates obtained by TRPL. Similar to MAPbI$_3$, the sum of mobilities of free charge carriers, including electrons and holes, increases upon cooling, then drops at 150 K, which is close to the phase change temperature point of FAPbI$_3$.[20,24] This turning point indicates that the onset of Auger recombination at 150 K and lower arises from a combination of both reduced carrier trapping efficiency and mobility drop.

With these preparations, we apply the temperature-dependent analyses to the FAPbI$_3$ EDTA-24h sample. All recorded temperature-dependent TRPL traces of FAPbI$_3$ EDTA-24h are shown in full in Fig. S3 and a subset of spectra shown in Fig. 2a-c. At lower temperatures, we observe increasingly pronounced curvature of the TRPL traces across the whole range under high excitation densities, which is the excitation conditions in which second order recombination dominates. Such a higher curvature indicates a larger value of $k_2$ at lower temperatures, and this is in line with previous temperature dependent records and analysis of charge carrier dynamics, such as THz[20], TRMC[24] and TRPL[25]. As shown in Fig. S3, the global fitting method performs well across a range of parameters, accurately capturing decays with varying curvatures at different temperatures. For each temperature, the fitting was performed separately via the global fitting by Equation 2. Additionally, the slope of the straight line in the log-linear plots, corresponding to first-order recombination, is smaller, suggesting a smaller value of $k_1$. These $k_1$ values are again double-checked (see Table. S1) by exponentially fitting the decay excluding the initial sharp drop, as shown in each inset in Fig. 2a-c. Then, as with the room temperature data, we use these $k_1$ values to perform PLQE analysis at each temperature (Fig. 2d-f). We find that the $k_2$ values obtained from TRPL by Equation 2 and PLQE by Equation 3 are consistent at each temperature. Based on the extracted $k_2$ values with good fits across all temperatures, we further find that $k_{2non}$ values increases from $2.75 \times 10^{-11}$ cm$^3$s$^{-1}$ to $6.8 \times 10^{-11}$ cm$^3$s$^{-1}$ and $k_{2non}$ / $k_2$ ratio drops from 98.9% at to 92.9%.

We apply our fitting approach combining TRPL and PLQE on various FAPbI$_3$ samples, as shown in Fig. S3-4 and summarized in Table S4. These include: (1) EDTA-std, a sample prepared using EDTA-doped precursors and surface-modified by gas quenching (standard method),[18] which exhibits stable quality and has a $k_{2non}$ of $9.75 \times 10^{-11}$ cm$^3$s$^{-1}$ ($k_{2non}$ / $k_2$ = 97.5%); (2) EDTA-24h, the sample shown above in Fig. 1 and 2, treated with EDTA and post-annealed in air for 24 hours at room atmosphere, which exhibits stable quality and an extended PL lifetime compared to the ETDA-std sample[18] (3) MACl, prepared using an FAPbI$_3$ precursor treated with MACl and surface modification,[15] resulting in an improved polycrystalline morphology that enhances charge collection, achieving a solar cell efficiency of 23% with a much lower $k_{2non}$ of 2.42×10$^{-11}$ cm$^3$s$^{-1}$ and a $k_{2non}$ / $k_2$ ratio of 75.6%; and (4) ACCl, a sample doped with ACCl and further surface-modified using CHABr and M-PEAI in IPA solution,[26] resulting in a uniform, pinhole-free surface with large grains. This material attains the highest solar cell efficiency of 25.1%, with a $k_{2non}$ of $2.68 \times 10^{-11}$ cm$^3$s$^{-1}$ and a $k_{2non}$ / $k_2$ ratio of 76.7%.

Despite the high qualities of either stability[18] or high device performance[15,26], all samples exhibit significant $k_{2non}$ values comparable to $k_2$ at room temperature. Previous work[8] using the combination of TRMC and PLQE analysis on perovskites verified that the overall $k_2$ can be affected by light-soaking post-treatment, highlighting the $k_{2non}$ component, as $k_{2r}$ is not expected to change since it is an intrinsic property. Here, we find different $k_{2non}$ proportions in the different FAPbI$_3$ samples at room temperature, which are as large as around 98% ($k_{2non}$ = 9.75 × 10$^{-11}$ cm$^3$s$^{-1}$) in EDTA samples and still significant at 77% in ACCl ($k_{2non}$ = 2.68 × 10$^{-11}$ cm$^3$s$^{-1}$), which shows state-of-the-art solar cell performance. After cooling, as shown in Fig.S7, the $k_2$ values of all samples extracted from TRPL and PLQE are consistent at each temperature, allowing us to analyse the temperature-dependent trend of $k_{2non}$ and $k_1$ values. As exhibited in Fig. 3 (a&c) and Fig. S6-8, which show the quantified temperature dependence of recombination rates, the trend of $k_{2non}$ values on temperature does not follow either $k_1$ or $k_2$. This discrepancy means there is no straightforward relationship between $k_{2non}$ and $k_1$ or mobility.

Firstly, the $k_1$ values decrease as temperature decreases for all samples (Fig. 1a), in line with a previous study[20] on halide perovskites and conforming to the standard behaviour of traditional semiconductors, such as recombination rates governed by deep traps[27]. The $k_1$ is traditionally described by a Shockley-Read-Hall model, and its decrease at low temperatures is attributed to reduced thermal energy, which diminishes impurity ionization, passivates dopant sites, and limits their role in recombination.[20] Meanwhile, the absolute $k_1$ value depends on the density and energy level position of traps. Therefore, due to the differences in doping and surfaces between the four samples, we observe a significant difference between the $k_1$ values as seen in Fig. 3a. Both MACl (7.7 × 10$^5$ s$^{-1}$) and ACCl (7.9 × 10$^5$ s$^{-1}$) samples, which are proven excellent solar cells, present significantly smaller $k_1$ than EDTA-std (7.8 × 10$^6$ s$^{-1}$). This suggests that bulk-doping modified film crystallisation leads to fewer deep traps in the materials. The improved[15,26] surface effect is supported by the EDTA-24h, where the 24-hour post-annealing in air primarily modifies the surface condition of the material compared to EDTA-std[18]. In this case, the $k_1$ of the EDTA sample drops dramatically (from 7.8 × 10$^6$ s$^{-1}$ to 5.2 × 10$^5$ s$^{-1}$) and becomes comparable to that of MACl and ACCl.

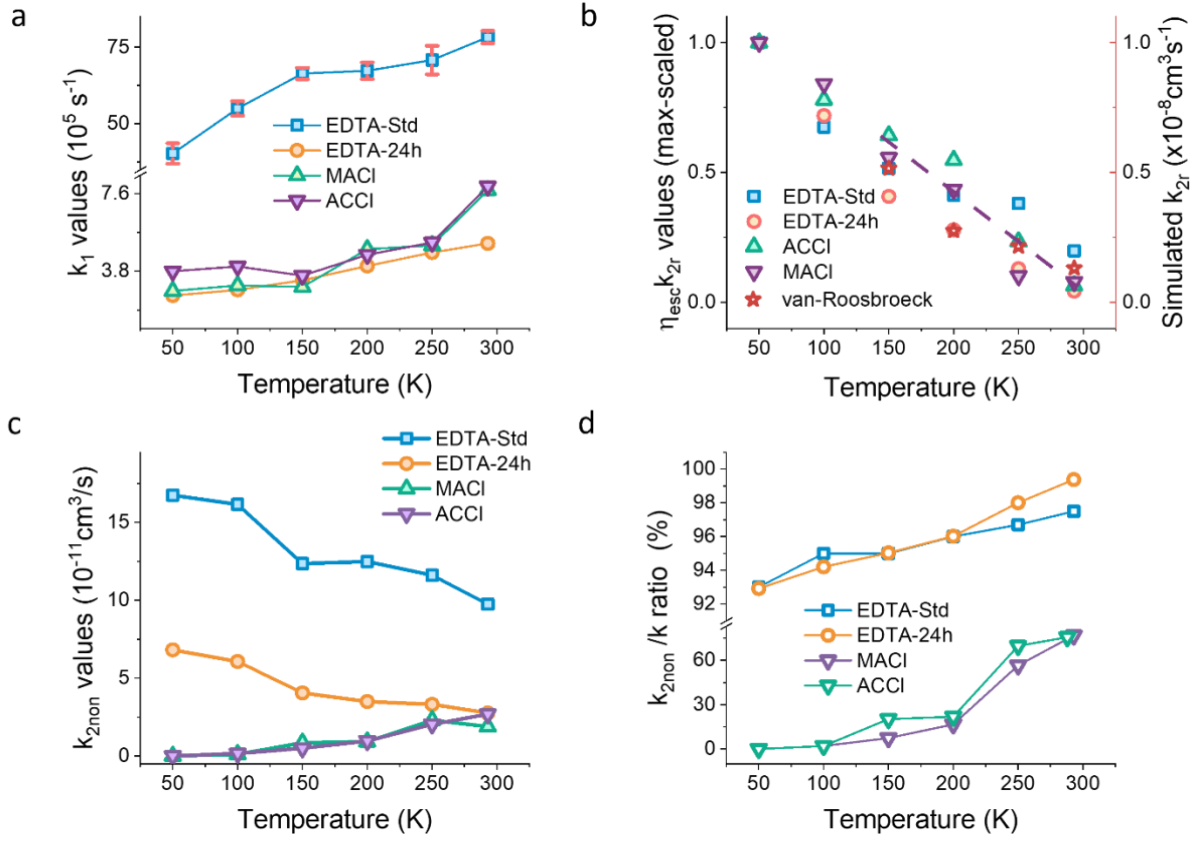

*FIG.3. Temperature-dependent recombination rates, extracted from fits to PLQE and TRPL data in Figures 2 and S3&4, using Equations 2 and 3, demonstrate the intrinsic and extrinsic properties of $k_2$. (a) $k_1$ values from TRPL fits, checked by exponential fits to the traces at the lowest excitation density. Error bars are representatively included for the EDTA-std sample. The error bar represents the standard error derived from the global fit covariance matrix, accounting for residual variance, data variability, and parameter correlations. Direct comparisons of $\eta_{esc}k_{2r}$ (b), $k_{2non}$ (c) and $k_{2non}/k_2$ ratio (d) between different samples are derived from the approach in section A. Normalized temperature-dependent $\eta_{esc}k_{2r}$ values with a dashed line (guide to the eye) illustrating their parallel nature to simulated $k_{2r}$ derived from the van Roosbroeck-Shockley relation.*

Secondly, the $k_2$ values increase as the samples are cooled, in contrast to $k_1$, as shown in Fig. S7 for comparison. This increase aligns with previous experimental reports about $k_2$ in halide perovskites, using various time-resolved spectroscopies[20,21]. It also follows the trend of temperature-dependent theoretical calculations on $k_2$.[21] In Fig. S6, we plot $\eta_{esc}k_{2r}$ values of each sample together with the simulated $k_{2r}$ via the van Roosbroeck-Shockley Relation, using temperature-dependent absorption (Supplementary Note 2). Importantly, we find that the lines of $\eta_{esc}k_{2r}$ values for each sample as a function of temperature are exactly parallel to the line of the simulated $k_{2r}$. The vertical offsets between these parallel lines arise exclusively from variations in $\eta_{esc}$, a factor critically influenced by surface treatment that differ among the four samples due to fabrication parameters, and also the difference in thickness.[28]

To further explore the parallel nature of these lines, we independently normalized the extracted $\eta_{esc}k_{2r}$ values to their respective maxima (Fig. 3b). This normalization shifts the curves vertically but preserves their temperature-dependent slopes, which were individually shown for all samples in Fig. S6. After scaling, the $\eta_{esc}k_{2r}$ values of the four samples—despite stark differences in doping, fabrication, and surface treatments—align onto the same line. This alignment reinforces that the temperature-dependent $k_{2r}$, extracted using Equation 3, is an intrinsic property of FAPbI$_3$, while the pre-normalization vertical offsets exclusively reflect extrinsic, sample-dependent variations in $\eta_{esc}$. This follows the principle of detailed balance, which states that the radiation by recombination and generation by absorption of electron-hole pairs are equal at thermal equilibrium. This means that the intrinsic second-order recombination is the reverse process of absorption of charge carriers, as has been reported for MAPbI$_3$.[21] According to the Van Roosbroeck-Shockley Relation, the increase in $k_{2r}$ during cooling can be attributed to the narrower Bose-Einstein distribution of electrons and holes, which increases the electron-hole interaction probability.[21]

Lastly, as shown in Fig. 3c, the trend of temperature-dependent $k_{2non}$ is not clearly following either $k_2$ or $k_1$. By contrast, the $k_{2non}/k_2$ ratio is monotonically decreasing over temperature for all samples (Fig. 3d), regardless of the significant discrepancy in values. MACl and ACCl, which have the smallest $k_1$ and $k_{2non}$, also show a smaller $k_{2non}/k_2$ ratio, dropping from around 75% to nearly negligible when cooling to 50 K. This suggests that, in these samples, the increase in $k_2$ as the sample is taken to lower temperatures is mainly due to $k_{2r}$. Despite a clear decrease, both the EDTA-std and EDTA-24h samples maintain a large ratio of about 93% of $k_{2non}/k_2$ at 50 K. This implies that at 50 K, trap-assisted first-order recombination is the only carrier loss mechanism in MACl and ACCl, while $k_{2non}$ still plays a significant role in the EDTA samples. Compared to the EDTA-std sample, the surface-modified EDTA-24h (post-treated) exhibits $k_1$ and $k_2$ values comparable to those of MACl and ACCl at room temperature. However, unlike MACl and ACCl, the $k_{2non}$ of EDTA-24h still increases upon cooling like EDTA-std. This means that in the case of EDTA-24h, of which the surface treatment is incomplete,[18] $k_{2non}$ remains pronounced at room temperature and persists at lower temperatures. Meanwhile, since $k_1$ is already significantly depressed, this case of EDTA-24h reinforces that $k_1$ and $k_{2non}$ are different.

### C. DFT supports that $k_{2non}$ is related to surface shallow states.

Despite the same temperature-dependent trend of $k_1$ and $k_{2non}$ of MACl and ACCl, the value of $k_1$ only drops 2 times while the $k_{2non}$ reduces to be negligible with respect to the absolute rates of other contributing terms with complete surface treatments, suggesting no connection between $k_1$ and $k_{2non}$. The distinction could indicate that $k_{2non}$ arises from shallow states, unlike $k_1$ from deep traps. In shallow states, carriers are weakly localized, allowing them to escape back to the conduction or valence band via thermally activated processes[29]. This thermal localization enhances $k_{2non}$ at lower temperatures, as cooling reduces carrier escape rates and increases states occupancy. However, this enhancement diminishes or even reverses if fewer carriers can be trapped—such as when the energy distribution and density of shallow states are substantially reduced. This is evidenced by Fig. 3c: EDTA-std and EDTA-24h samples exhibit elevated $k_{2non}$ at lower temperatures, while MACl and ACCl display suppressed $k_{2non}$ values, owing to effective surface modification for the latter samples.

In weakly trapping shallow states, electron-hole recombination resembles band-to-band processes in its dependence on both the concentration of electrons and holes but becomes non-radiative due to the localization of one carrier. This second-order effect can be considered as an extreme case of the Shockley-Read-Hall recombination model with shallow traps. Our experimental results support this model over trap-assisted Auger recombination (TAAR), which for example leads to $k_{2non}$-type processes in III-nitride semiconductors.[30] TAAR is a pseudo-second order process in which a trapped carrier is involved in the third-order Auger recombination. However, due to the trap density overwhelming the other two, the trapped concentration in the recombination form can be scaled into the coefficient, leading to a pseudo-second order expression. Representative cases of the TAAR process include a highly doped semiconductor such as silicon[31], or semiconductor nanocrystals[32]. The key common factors that lead to TAAR in these cases are high carrier density (from heavy doping or strong confinement) and high trap concentration (from doping or a large surface area to volume ratio). In our case, Auger recombination only appears at low temperatures in the recorded excitation density range (or at low temperature, the onset of Auger recombination shows up at lower densities), consistent with reduced carrier trapping and the change in mobility spike, but $k_{2non}/k_2$ is larger at room temperature, in contrast to the temperature-dependence of TAAR[33]. Also, the carrier density after excitation in our work is comparable to 1 sun illumination, which does not yield a high carrier concentration[34]. Moreover, the opposite trend of temperature-dependent $k_1$ and $k_{2non}$ in both EDTA-std and EDTA-24h suggests that the deep trap concentration is not directly involved in $k_{2non}$ processes.

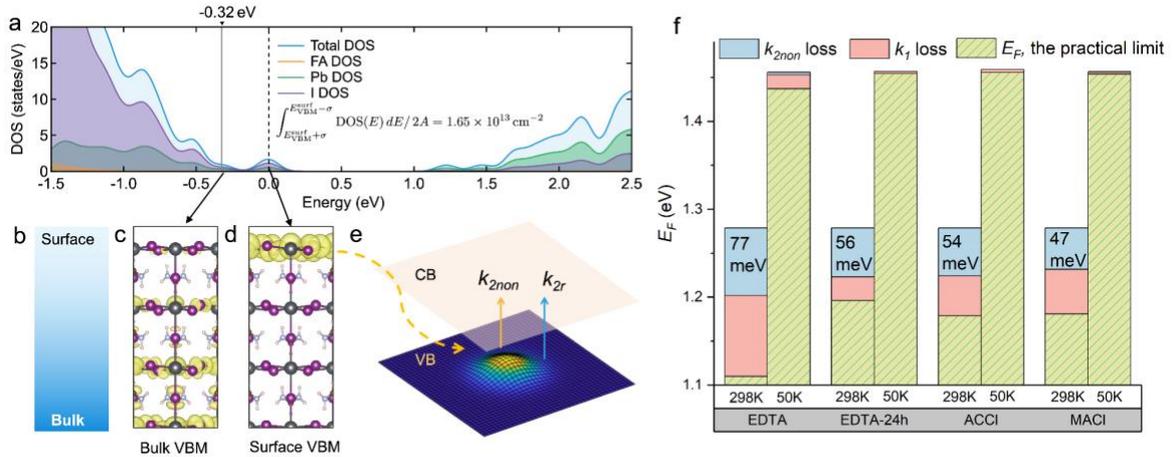

FIG. 4. (a-d) DFT calculations of PbI$_2$ termination at the surface. (b) shows the positions of calculated density of states from bulk towards surface for (c) and (d). As illustrated in (d), Shallow states adjacent to the valence band, extending to a DFT-calculated reference value of ~320 meV, are generated by the surface termination of PbI$_2$. These surface shallow states act as a 'collection bowl' for holes, thus leading to the $k_{2non}$ process as depicted in (e), along with $k_{2r}$. (f) Calculated $E_F$ at room temperature and 50K. The $k_{2non}$ loss contributes 47-77 meV of the overall carrier losses at ambient condition and nearly negligible at 50K. The bar cap

*represents the detailed balance limit*[35]*, and the $k_{2non}$ or $k_1$ loss is calculated by including or setting zero the $k_{2non}$ or $k_1$ values while determining the quasi-Fermi level splitting, $E_F$.*

A relatively high concentration of trap density is possible at the surface or grain boundaries where the density of states can differ from the bulk due to lattice termination. The $k_{2non}$ values of the samples, labelled as EDTA-std for the bare film and EDTA-24h for film with surface modified after treatment, present a significant difference in Fig. 3c, supporting the relevance of the surface. This is further supported by the other two samples, ACCl and MACl, both of which show much smaller $k_{2non}$ with specific surface modification. The possibility of relatively concentrated surface traps also aligns with wider literature[8,36,37] consistent with $k_{2non}$ primarily being a surface effect. The light soaking effect, which enhanced PL lifetime and PLQE, was originally demonstrated in MAPbI$_3$ and was proposed to be effective on surface shallow states[8]. The same treatment on FAPbI$_3$ also led to better PL performance but, meanwhile, bulk properties were not affected.[18]

To understand such an effect, we conducted DFT calculations that suggest a potential cause for $k_{2non}$, as shown in Fig. 4a-d. We find that in the case of PbI$_2$ surface terminations, [PbI$_6$]$^{4-}$ octahedra are truncated at the surface, thus creating an extra projected density of states to be $1.6 \times 10^{13}$ cm$^{-2}$ at the surface, spanned by a reference value of ~320 meV adjacent to the valence band. By contrast, for FAI-terminated surfaces, the [PbI$_6$]$^{4-}$ octahedra remain structurally intact with no truncation, as shown in Fig. S9. Consequently, no such extra projected density of states at the surface are observed in this configuration. Such a result resembles the energy levels of the surface states that are located just above the top of the valence band in bulk MAPbI$_3$.[35] Holes could accumulate in these shallow states while electrons will remain in the bulk region.[36] One possible mechanism for $k_{2non}$ is thus illustrated in Fig. 4e. These locally accumulated holes, which hop between these surface states and bulk states, involve the overall carrier density in the valence band in their recombination, leading to a second-order recombination behaviour but with a rate different from band-to-band. The shallow states responsible for $k_{2non}$, which arise intrinsically from the surface termination of PbI$_2$, support the ubiquitous presence of $k_{2non}$ in lead halide perovskites[13]. Such surface termination-induced shallow states are in line with the experimental results that the surface-targeted treatment, either gas quenching or post-treatment, could effectively decrease $k_{2non}$.

Given the shallow states are from the PbI$_2$ surface terminations, these states are localised and energetically distinct from bulk FAPbI$_3$ bands. At lower temperatures, some surface shallow states (on the deeper side) behave like deep traps because cooling reduces the thermal energy required for holes to escape back to the valence band. These states now act as deep hole traps, forcing recombination to proceed same as $k_1$. This reduces the $k_{2non}/k_2$ ratio because fewer active shallow traps contribute to $k_{2non}$, meanwhile $k_{2r}$ increases[21]. Although $k_{2non}/k_2$ ratio decreases at lower temperatures for all samples, the absolute $k_{2non}$ values exhibit distinct trends: they increase for EDTA-std and EDTA-24h but decrease for surface-modified samples (MACl and ACCl). Notably, the $k_{2non}$ reduction in surface-modified samples follows an exponential temperature dependence, consistent with an activation energy process. This behavior is corroborated by the Arrhenius plot in Fig. S10, which reveals an activation energy of ~40 meV.

The presence of this energy barrier indicates that surface modification suppresses carrier recombination by introducing a kinetic hindrance. By contrast, EDTA-std (non-passivated) and EDTA-24h (partially modified) lack such an activation barrier. Their elevated $k_{2non}$ at lower temperatures instead is solely dominated by thermal localization effects (reduced carrier escape, higher state occupancy). Given that PbI$_2$ terminations can be tailored via synthesis or post-treatment, these results highlight a viable pathway to control or even eliminate $k_{2non}$ by engineering surface terminations.

### D. The quantified effect of $k_{2non}$ on the open-circuit voltage of perovskite solar cells

Finally, we evaluate the effect of $k_{2non}$ on the practical performance of solar cells. In Fig. 4c, we calculate the quasi-Fermi level splitting (i.e., difference in energy between the electron and hole populations after photoexcitation), $E_F$, calculated using the recombination rates and generation term from 1 sun illumination at AM1.5 (Supplementary Note 3). Since $E_F$ directly links to the open circuit voltage, $V_{oc}$, $E_F$ can be taken as the practical limit of $V_{oc}$, as previously evaluated for perovskite solar cells.[2] As noted in the room temperature column for each sample, the loss due to $k_{2non}$ contributes to as low as 47 meV (MACl) up to 77 meV (EDTA) of the total carrier losses. Even in state-of-the-art devices such as ACCl (achieving a PCE of 25.1%), a residual energy loss of 54 meV persists. This underscores the need for targeted mitigation strategies from $k_{2non}$, despite significant breakthroughs in film quality and device optimization.

In line with the observation in Fig. 3 that $k_{2non}$ and $k_1$ are not directly correlated, the losses associated with $k_1$ and $k_{2non}$ differ in proportion, as shown in Fig. 4f. The distinct $E_F$ drop caused by $k_1$ (1st-order, via deep traps) and $k_{2non}$ (2nd-order, via shallow traps) underscores the two separate populations of traps and their fundamentally different recombination physics in solar cells. Both need addressing, and targeting elimination of these losses is critical for device optimization. The observed reduction in the $E_F$ caused by $k_{2non}$—when comparing the EDTA-std to the three other cases—underscores the potential of targeting surface conditions to mitigate $k_{2non}$ losses and progress toward the radiative limits. Since dopants can interact with PbI$_2$ during crystallization of FAPbI$_3$[26]—resulting in a smoother, more uniform surface—and the DFT simulation suggests that PbI$_2$ termination generates the shallow states, targeting terminations during and after crystallisation presents a promising approach to further reducing $k_{2non}$. Moreover, we also evaluate $E_F$ at 50 K with the parameter values obtained above. From the negligible effect of $k_{2non}$ loss along with the decrease in $k_1$ loss upon cooling, we expect a much higher $V_{oc}$, approaching the theoretical limit, and thus better device performance under cooler conditions, such as those implemented in space applications.

### III. CONCLUSIONS

We have elucidated the origins and impacts of the second-order non-radiative recombination pathway ($k_{2non}$) in halide perovskites. By combining fluence- and temperature-dependent TRPL and PLQE, and manipulating charge carrier dynamics through control of bulk and surface conditions, we reveal the radiative component ($\eta_{esc}k_{2r}$) of second-order recombination aligns with the simulated $k_{2r}$ values from the Van Roosbroeck -Shockley Relation as a function of temperature. We demonstrate that $k_{2non}$ arises from shallow states, distinct from the first-order $k_1$ pathway mediated by deep traps. Both recombination pathways of carrier losses need to be addressed in device optimization strategies. Our analysis reveals that these shallow states are primarily governed by extrinsic factors, with PbI$_2$ terminations identified as a likely contributing factor through DFT calculations. Our further calculations show that the current gap between the best achieved $V_{oc}$ to date and the limit, due to $k_{2non}$ loss, is up to ~80 mV. This work not only enhances our understanding of $k_{2non}$ but also provides actionable directions, emphasizing surface conditions to mitigate its effects, thereby paving the way for further improvements in perovskite solar cell efficiencies. Moreover, given $k_{2non}$ loss is critical to address in single-junction cells, its impact remains equally vital when these wide gap cells are integrated into tandem architectures, as second-order non-radiative recombination directly weakens the luminescent coupling between sub-cells.[38,39]

## ACKNOWLEDGMENTS

The authors thank

European Union's Horizon 2020 Research and Innovation Program, European Research Council, HYPERION, 756962 and PEROVSCI, 957513 (SDS, LD)

EPSRC EP/V012932/1 and (SDS, SN)

Sir Henry Royce Institute grant EP/R00661X/1 and EP/P024947/1 (SDS)

CAM-IES grant EP/P007767/1 (SDS)

European Research Council under the European Union's Horizon 2020 research and innovation programme Grant Agreement No. SCORS – 101020167 (SG)

Royal Society and Tata Group (grant no. UF150033, URF\R\221026) (SDS)

Engineering and Physical Sciences Research Council EP/V06164X/1(LY, SDS)

UKRI guarantee funding for Marie Skłodowska-Curie Actions Postdoctoral Fellowships 2021 (EP/X025756/1) (YKJ)

## AUTHOR DECLARATIONS

### Conflict of Interest

Samuel D. Stranks is a co-founder of Swift Solar Inc.

### Author Contributions

Dengyang Guo: Conceptualization (lead); Data curation (lead); Investigation (lead); Methodology (equal); Software (equal); Validation (equal); Visualization (lead); Writing – original draft (lead); Writing – review & editing (equal). Alan R. Bowman: Formal analysis (equal); Methodology (equal); Software (equal); Validation (equal); Writing – review & editing (equal). Sebastian Gorgon: Formal analysis (equal); Data curation (equal); Writing – review & editing (equal). Changsoon Cho: Formal analysis (equal); Software (equal); Writing – review & editing (equal). Youngkwang Jung: Formal analysis (equal); Software (equal); Data curation (equal); Writing – review & editing (equal). Jiashang Zhao: Formal analysis (equal); Validation (equal); Data curation (equal); Writing – review & editing (equal). Linjie Dai: Formal analysis (equal); Data curation (equal); Writing – review & editing (equal). Jaewang Park: Data curation (equal); Writing – review & editing (equal). Kyung Mun Yeom: Data curation (equal); Writing – review & editing (equal). Satyawan Nagane: Data curation (equal); Writing – review & editing (equal). Stuart Macpherson: Data curation (equal); Writing – review & editing (equal). Weidong Xu: Data curation (equal); Writing – review & editing (equal). Jun Hong Noh: Formal analysis (equal); Funding acquisition (equal); Project administration (equal); Resources (equal); Supervision (equal); Visualization (equal); Writing – review & editing (equal). Sang Il Seok: Formal analysis (equal); Funding acquisition (equal); Project administration (equal); Resources

(equal); Supervision (equal); Visualization (equal); Writing – review & editing (equal). Tom Savenije: Formal analysis (equal); Funding acquisition (equal); Project administration (equal); Resources (equal); Supervision (equal); Visualization (equal); Writing – review & editing (equal). Samuel D. Stranks: Investigation (equal); Formal analysis (equal); Funding acquisition (lead); Methodology (equal); Project administration (lead); Resources (lead); Software (equal); Validation (equal); Visualization (equal); Writing – review & editing (equal).

## DATA AVAILABILITY

The data supporting the findings of this study are available from the corresponding author upon request.

**Supplementary Information**

**Title:** Competition Between Controllable Non-Radiative and Intrinsic Radiative Second-Order Recombination in Halide Perovskites

# Table of Contents

**Materials**
EDTA-std
EDTA-24h
ACCl
MACl

**Methods**
PLQE
TRPL
TRMC
UV-VIS

**Supplementary Tables**
TABLE S1 The double-checked $k_1$ values, one from the global fitting, and the other from single exponential fitting on the trace obtained from the lowest excitation density.
TABLE S2 The parameters obtained from the combined fitting approach
TABLE S3 The dynamic rates obtained from exponential fits
TABLE S4. The sample information.

**Supplementary notes**
supplementary Note 1. The bi-exponential fitting is inappropriate for carrier dynamics analysis
supplementary Note 2. $k_2$ simulation by van-Roosbroeck and Shockley Relation
supplementary Note 3. quasi-Fermi-level splitting calculations

**Supplementary Figures**
FIG.S1 TRPLs of the four samples at a high (A) and (B) low excitation density.
FIG.S2 Temperature dependent PL spectra of samples EDTA-std, MACl, and ACCl.
FIG.S3 Temperature dependent PL decays and fits of each sample
FIG.S4 Temperature dependent PLQE and fits of each sample
FIG.S5 Temperature dependent TRMC (normalised to one) of EDTA-24h
FIG.S6 Temperature dependent $k_1$ values along with the sum of mobilities
FIG.S7 Trend agreement between the $k_2$ values extracted from TRPL and PLQE at each temperature
FIG.S8 Temperature dependent $k_1$ and $k_2$ values of each sample.
FIG.S9. DFT calculations of ots-FAPbI$_3$ (001) surface.
FIG.S10. Arrhenius plot of the natural logarithm of MACl and ACCl versus inverse temperature (1/T).

**Materials**

**EDTA-std films**[1]

Materials: The following reagents were sourced from commercial suppliers and employed without additional purification: N,N-dimethylformamide (DMF, anhydrous, 99.8%), dimethyl sulfoxide (DMSO, anhydrous, 99.9%), isopropanol (anhydrous, 99.9%), ethylenediaminetetraacetic acid (EDTA, ≥99%), and chlorobenzene (CB, 99.99%) from Sigma-Aldrich. Formamidinium iodide (FAI, 99.9%) was procured from Greatcell Solar Materials. Lead iodide ($PbI_2$, 98%) awas supplied by TCI. All materials were utilized in their as-received form.

Sample fabrication: A precursor solution was prepared by dissolving lead iodide ($PbI_2$, 0.346 g, 1.5 mmol), formamidinium iodide (FAI, 0.155 g, 1.8 mmol), and ethylenediaminetetraacetic acid (EDTA, 5 mol% relative to $PbI_2$) in 0.5 mL dimethyl sulfoxide (DMSO). The mixture was stirred continuously at 75°C until a homogeneous dark yellow solution formed. This solution was deposited onto a UV-ozone-treated glass substrate via spin-coating (4,000 rpm, 40 seconds). To enhance film uniformity, a gas-quenching step was implemented during the spin-coating process: nitrogen gas was directed onto the rotating substrate starting 5 seconds after spin initiation, maintaining a flow duration of 20 seconds. The nitrogen delivery nozzle was initially positioned ~7 cm from the film surface for the first 10 seconds, then adjusted to ~5 cm for the remaining 10 seconds. Post-deposition, the films underwent thermal annealing at 150°C for 1 hour within a nitrogen-filled glovebox.

**EDTA-24 films**[1]

The EDTA-24h films were produced by annealing EDTA-std films in a room atmosphere at 100 °C for 24 hours.

**MACl films**[2]

The MACl sample was synthesized using a spin-coating technique involving a two-step procedure (5 seconds at 1,000 rpm followed by 15 seconds at 5,000 rpm). The precursor solution consisted of formamidinium iodide (FAI, 1.6 M), lead iodide ($PbI_2$, 1.6 M), and methylammonium chloride (MACl, 0.45 M) dissolved in a solvent mixture containing 0.8 mL dimethylformamide (DMF) and 0.1 mL dimethyl sulfoxide (DMSO). This solution was deposited onto a fused silica substrate. During the final stage of spinning, 1 mL of ethyl ether was rapidly dispensed onto the substrate. Subsequent annealing at 150°C for 10 minutes induced crystallization of the film while facilitating the evaporation of MACl.

**ACCl films**[3]

Materials: The following materials were used in the experiments formamidinium iodide (FAI; 99.99%, Greatcell solar), lead(II) iodide ($PbI_2$; 99.99%, TCI), methylammonium chloride (MACl; Merck), propylamine hydrochloride (PACl; Sigma-Aldrich), Acetylcholine chloride(ACCl; 99%, Sigma-Aldrich), Cyclohexylammonium bromide ((CHA)Br; Great

Solar), 4-methoxy-phenethylammonium iodide (MeO-PEAI; 99%, Greatcell Solar), N,N-dimethylformamide (DMF; 99.8%, Sigma-Aldrich), dimethyl sulfoxide (DMSO; 99.9%, Sigma-Aldrich), 2-propanol (99.5%, Sigma-Aldrich).

Sample fabrication: Glass substrates were sequentially sonicated in acetone, detergent, and ethanol for 20 min each. The perovskite precursor solutions were prepared by dissolving 1.4 M $FAPbI_3$ powder with ACCl (0.75 mol%), PACl (4mol%) and MACl (35 mol%) in a mixed solvent of DMF: DMSO = 8:1. The perovskite solutions were then spin-coated onto the substrate at 1,000 rpm for 10 s and 5,000 rpm for 15 s; 600 ul of ethyl ether was dripped onto the substrate during spinning. All perovskite layers were heat-treated at 120 °C for 1 h. After cooling, CHABr and M-PEAI (8 mM and 16 mM, respectively) in IPA solution was spin-coated on the perovskite-coated substrate at 5,000 rpm for 30 s and annealed at 100 °C for 5 min.

**Methods**

**PLQE**

The photoluminescence quantum efficiency (PLQE) was determined using an integrating sphere, adopting the three-step measurement protocol established by DeMello et al.[4] A temperature-stabilized 520 nm continuous-wave laser (Thorlabs) served as the excitation source, with its fluence adjusted via an optical filter wheel. Emitted light was captured using a high-sensitivity silicon detector (Andor iDus DU420A). To maintain measurement consistency, the integrating sphere underwent periodic cleaning and recoating, and system-specific calibration procedures were performed prior to data collection.

**TRPL**

Time-resolved photoluminescence (TRPL) was recorded by iCCD, a gated intensified CCD detector (Andor iStar DH740 CCI-010) coupled to a calibrated spectrograph (Andor SR303i) that captured the PL signals. Excitation was provided by frequency-doubling the 800 nm output of a Ti:sapphire optical amplifier (1 kHz repetition rate, 90 fs pulse duration) and BBO to achieve narrowband 520 nm irradiation. The incident excitation density varied from $2\times 10^{15\text{-}17}$ $cm^{-3}$. These initial excited carrier densities were calculated according to the method of Richter et al.[5] The effective beam diameter of the excitation spot was 990 μm. For the temperature-dependent TRPL measurements, the sample was secured within a helium-cooled closed-cycle cryostat (Optistat Dry BL4, Oxford Instruments) located at the intersection of the beam paths. The cryostat system was regulated by a compressor unit (HC-4E2, Sumitomo) integrated with a digital temperature controller (Mercury iTC, Oxford Instruments). During experiments, the internal environment of the cryostat was maintained under a vacuum pressure lower than $10^{-5}$ mbar.

## TRMC

The time-resolved microwave conductivity (TRMC)[6] method measures transient changes in microwave power reflection from a sample-loaded cavity following pulsed laser excitation. Photoconductance ($\Delta G$) is calculated using the relationship between the normalized microwave power decrease ($\Delta P/P$) and the sensitivity factor $K$:

$$-K \cdot \Delta G(t) = \Delta P(t)/P$$

The yield of charge carrier ($\varphi$) and effective mobility sum ($\sum \mu = \mu e + \mu h$) were determined via:

$$\varphi \sum \mu = \frac{\Delta G}{I_0 \beta e F_A}$$

Here, $I_0$ represents the incident photon flux per pulse per unit area, $\beta$ is the microwave cavity's geometry constant, $e$ is the elementary charge, and $F_A$ denotes the fraction of absorbed light at the 500 nm excitation wavelength ($F_A$=0.8). ow-temperature TRMC measurements were conducted by cooling the samples with liquid nitrogen. Once the target temperature was reached, the samples were allowed to stabilize for at least 15 minutes to ensure thermal equilibrium.

## UV-VIS

UV-VIS absorption measurements under variable temperature conditions were conducted using a dual-beam spectrophotometer (Cary 6000i, Agilent Technologies). The sample was housed in an optically accessible cryostat (Optistat CF-V, Oxford Instruments) fitted with quartz windows. Experiments were performed in transmission mode under an ultra-high vacuum environment at approximately $10^{-6}$ mbar.

## Supplementary Tables

TABLE S1 The double-checked $k_1$ values, one from the global fitting, and the other from single exponential fitting on the trace obtained from the lowest excitation density.

| Temperature | 293K | 250K | 200K | 150K | 100K | 50K |
| --- | --- | --- | --- | --- | --- | --- |
| $k_1$ (s$^{-1}$) from global fitting | 516398 | 471886 | 405921 | 336034 | 381551 | 260934 |
| $k_1$ (s$^{-1}$) from single fitting | 460984 | 456319 | 43417 | 409515 | 348649 | 303808 |

TABLE S2 The parameters obtained from the PLQE fitting approach

| Sample | $k_2$ (cm$^3$s$^{-1}$) | $k_1$ (s$^{-1}$) From TRPL | $k_{2non}$ (cm$^3$s$^{-1}$) | $\eta_{esc}k_{2r}$ (cm$^3$s$^{-1}$) | $k_{2non}/k_2$ ratio(%) |
|---|---|---|---|---|---|
| EDTA-24h | 2.78 ×10$^{-11}$ | 516398 | 2.75 ×10$^{-11}$ | 0.02 ×10$^{-11}$ | 98.9 |

TABLE S3 The lifetimes obtained from exponential fits

| $n_0$ (cm$^{-3}$) | Lifetime (ns) | EDTA | EDTA-24h | ACCl | MACl |
|---|---|---|---|---|---|
| ~ $1 \times 10^{17}$ | $\tau_1$ | 9.51 | 42.39 | 33.65 | 33.98 |
|  | $\tau_2$ | 27.60 | 170.57 | 138.17 | 192.56 |
| ~ $1 \times 10^{15}$ | $\tau$ | 50.59 | 287.18 | 96.20 | 304.25 |

TABLE S4. The sample information. The samples studied in this paper show outstanding stability[1] or Power Conversion Efficiency (PCE)[2,3], named as EDTA-std , EDTA-24h, MACl, and ACCl, respectively. All of them exhibit significant $k_{2non}$ values and high $k_{2non}/k_2$ at room temperature. Here, $k_2$ is the overall second order recombination rate, subject to $k_2 = \eta_{esc}k_{2r} + k_{2non}$

| Samples | Key Processing Steps [ref] | Quality | $k_{2non}$ (cm$^3$s$^{-1}$) $k_{2non}/k_2$ at room T |
|---|---|---|---|
| EDTA-std | precursor doped by EDTA, standard[16] | Stable | 9.75× 10$^{-11}$ 97.5% |
| EDTA-24h | precursor doped by EDTA, surface modified by post annealing treatment[16] | Stable+longer PL lifetime | 2.75× 10$^{-11}$ 98.9% |
| MACl | precursor treated by MACl and surface modification proved[13] | High PCE (23%) | 2.42× 10$^{-11}$ 75.6% |
| ACCl | precursor doped by ACCl and surface modified by CHABr and M-PEAI in IPA solution[25] | High PCE (25.1%) | 2.68× 10$^{-11}$ 76.7% |

## supplementary Note 1. The bi-exponential fitting is inappropriate for carrier dynamics analysis

First, we find that a commonly used method for analysing charge carrier dynamics in perovskite solar cell devices can only be used for comparison and does not provide true values of the dynamic parameters. Meanwhile, some methods based on time-resolved spectroscopies, while comprehensive, may require more complex analyses that are not ideal for quick preliminary checks in device fabrication. Fig. 1A and B show time-resolved Photoluminescence (PL) results (lines) and fitting (dashed lines) of the currently most efficient perovskite, formamidinium lead iodide ($FAPbI_3$). We study $FAPbI_3$ samples which either show outstanding stability[1] or device performance[2,3], named as EDTA-std, EDTA-24h for standard and 24h post-treated EDTA-stabilised $FAPbI_3$, and MACl and ACCl for each doped $FAPbI_3$ respectively. Importantly, these samples differ specifically by surface modification and bulk doping effect, which result in different charge carrier dynamics. As generally observed, the TRPLs under excitation density close to the standard solar illumination condition (so-called one sun at AM1.5) are curved in Fig 1A and become straight under a much lower excitation density in Fig.1B. This condition is pinpointed by the equivalent pulsed excited carrier density, under which carrier recombination mirrors the recombination observed under solar illumination. The change of decay is acknowledged as changing from second order (or bimolecular) recombination to first order (or mono-molecular) recombination of free charge carriers. Since TRPLs are recorded by pulsed excitations, we point out the solar illumination condition by the equivalent initial pulsed excited carrier density under which carrier recombination undergoes the same recombination under solar illumination.

We first address the issue of the prevalent use of biexponential fitting in dynamic analysis. We stress that this approach is inappropriate as it is not based on physical processes. To overcome the limitations, we introduce a simple approach that adheres to the physics process.The fits lines were obtained by the commonly used method: fitting by bi-exponential and exponential formulas, providing a numerical comparison of charge carrier lifetimes. As listed in table S3, the fastest TRPL from EDTA-std seen in Fig.1A and B has a lifetime of 9.51+27.60 ns under an excitation density of $1\times10^{17}$ $cm^{-3}$ and 50.59 ns under $1\times10^{15}$ $cm^{-3}$, while the longest trace from MACl has a lifetime of 42.39+170.57 ns and 304.25 ns, correspondingly.

We stress that although the lifetimes could numerically describe the fast and slow recombination processes under high excitation density from observation, the obtained parameters listed above cannot represent the actual opto-physical processes and are not suitable for recombination analysis. Regarding the recombination processes, the standard equation is:

$$\frac{dn}{dt} = G - k_1 n - k_2 n^2 - k_3 n^3 \qquad (S1.1)$$

Here, the $k_1$, $k_2$, and $k_3$ represent the first, second and third order (Auger) recombination rates, respectively. $G$ is the generation term, which is time-dependent for continuous excitation and time-independent for pulsed laser excitation. When the photogenerated carrier density, $n$ is much smaller than the background carrier concentration $p_0$ ($n \ll p_0$), the carrier recombination is dominated by the first order, and so the carrier density as a function of time can be integrated as $n = n_{init} exp(-k_1 t)$ ,[5] where $n_{init}$ stands for initial carrier density generated from excitation. In this case, $n$ is also much smaller than that generated from the standard solar illumination condition, so we call the excitation density low. Note, that for a specific sample, the definition of 'low density' depends on the actual decay whether it remains in the lowest order recombination regime, not necessarily less than one sun illumination. Physically, PL is a second-order recombination process involving both photoexcited electrons and holes: $PL = k_2 n^2$. Whereas, under low excitation densities,
$PL = k_2 n^2 = k_2(n_{init} exp(-k_1 t))^2 = k_2 n_{init}^2 exp(-2k_1 t)$, so such a second order recombination process is governed by a first order decay[5]. In this case,
PL can indeed be analysed by a mono-exponential formula $y = A exp(-t/\tau)$. Therefore, it shows as a straight line in a log-lin plot, as shown in Fig. 1B.

However, the commonly used bi-exponential formula
$y = A_1 exp(-x/\tau_1) + A_2 exp(-x/\tau_2)$
is a mathematically simple summation of two mono-exponential components. Even though such a summation can give a seemingly nice match between the fits and experimental records, physically it stands for an addition of two first-order recombination governed processes, indicating two different trap-dominated charge carrier recombination processes rather than second-order recombination. This is further emphasised by the differences in the values obtained in Table S3. Also, mathematically, the fitting is rather arbitrary and gives different values from each time's fitting. Therefore, it is incorrect to name the fast value from this bi-exponential fitting as second-order recombination lifetime. Also, under the excitation close to or larger than one sun illumination, trap-domination means severe loss of free charge carriers and is in contrast to the outstanding optoelectronic performance of perovskites in either solar cells or LEDs.

The ideal case for solar cell applications is described by the detailed balance in the well-known Shockley–Queisser limit, where there is only radiative second-order recombination which maintains temperature equilibrium with the environment via radiance. In this case, trap-assisted recombination is either non-existent or negligible, especially under one sun illumination. A direct effect of trap-assisted recombination on the optoelectronic property is the photoluminescence quantum efficiency (PLQE). As illustrated in Fig. 1C, these films show various PLQE values from different samples, ranging from 0.2% to 30%, under the excitation density close to one sun illumination, with a fixed excitation wavelength of 520 nm. We select 520 nm since at this excitation wavelength the absorbance is constant against temperature, as shown in Fig.1D, which will also facilitate a temperature-dependent analysis. Despite the absolute value difference, there is a turnover from linear to curved dependence of PLQE over excitation density, as indicated by the dashed lines. Such a turnover is in line with the dominance by first-order or trap-assisted recombination to second-order recombination.[7] The MACl and ACCl sample, which exhibits the highest reported solar cell efficiency to date for FAPbI$_3$[2,3], shows the smallest turnover, suggesting the least trap-assisted charge carrier recombination in FAPbI$_3$.

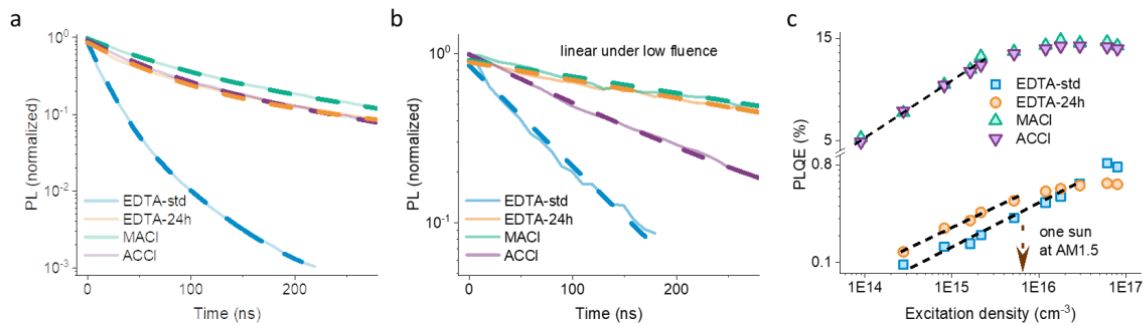

FIG.S1. TRPLs of the four samples at a high (a) and (b) low excitation density. The high and low densities are defined by the density generated from the standard one sun illumination. The seemingly good match between the data and bi-exponential fits in (a) is commonly used for $k_2$

discussion, which is incorrect in both physics and mathematics aspects. On the other hand, the mono-exponential fits in (b) are proper for extracting $k_1$. (c) Excitation density dependent PLQE of the four samples. The turning points correspond to the change from first order to second order carrier recombination.

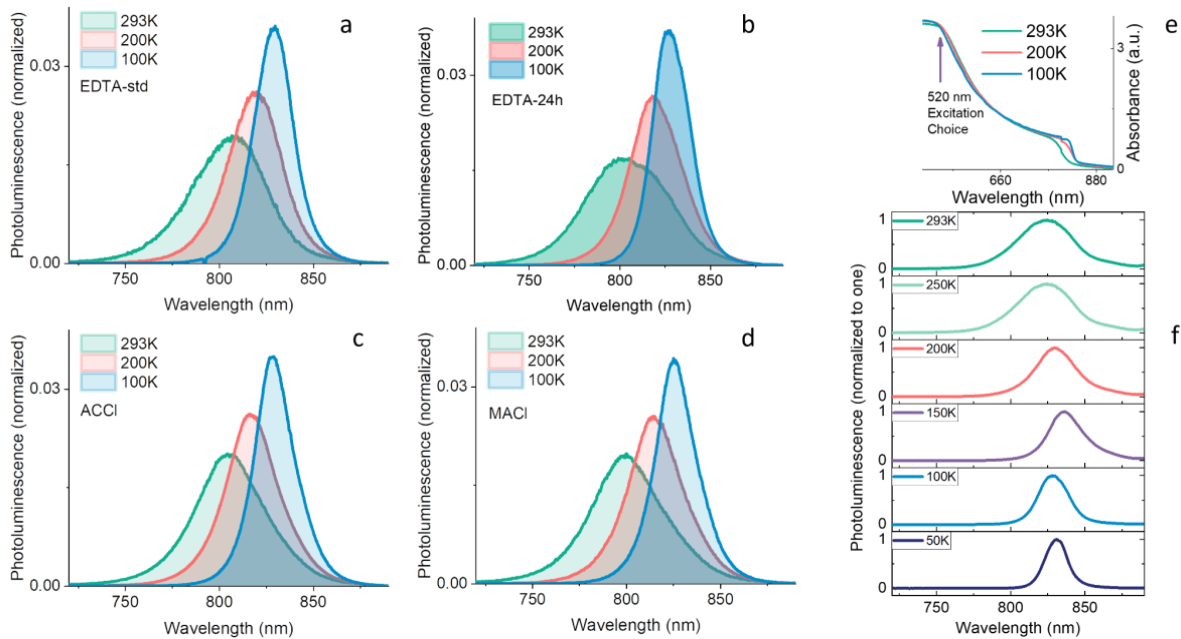

FIG.S2 Temperature-dependent PL spectra comparison between samples: DTA-std (a), EDTA-24h (b), ACCl (c), and MACl (d). These PLspectra were measured by steady-state excitation with a fluence of $2\mu J/cm^2$, and normalised by the integrated area over the emission range. (e) Temperature dependent absorbance of the sample EDTA-24h. The absorbance is constant at 520 nm while changing temperature, so this wavelength will be applied for temperature dependent measurements. (f) A full display of temperature dependent PL of EDTA-24h as an example, showing single peak emissions which are ideal for the PLQE scaling processing.

*supplementary Note 2 $k_2$ calculation by van-Roosbroeck and Shockley Relation*

**Way.1**
The $k_2$ coefficients were calculated using the van-Roosbroeck-Shockley Relation[8,9]. According to the detailed balance between the emission and absorption at each temperature ($T$), the intrinsic thermal radiation is:

$$R_{rad,i}(T) = k_{2r}(T)\, n_i(T)^2 = \int 4n^2\, \Phi_{BB}(\lambda,\, T)\, \alpha(\lambda,\, T) / E_{ph}(\lambda)\, d\lambda, \qquad (S2.1)$$

where $n$ is the refractive index of perovskites (assumed to 2.5 here), $\lambda$ is wavelength, $\alpha(\lambda,\, T)$ is the absorption coefficient spectrum, converted from absorbance measured at each temperature, $E_{ph}(\lambda)$ is the photon energy ($= hc/\lambda$; $h$ = Planck constant; $c$ = speed of light), and $\Phi_{BB}(\lambda,\, T)$ is the spectral photon flux of blackbody radiation at $T$, obtained by:

$$\Phi_{BB}(\lambda,\, T) = 2\, h\, c^2\, \lambda^{-5}\, (\exp(hc/\lambda k_B T) - 1)^{-1}, \qquad (S2.2)$$

with Boltzmann constant, $k_B$. Therefore, $k_{2r}$ can be obtained by $k_{2r}(T) = R_{rad,i}(T) \, n_i(T)^{-2}$ in equation S2.1, where

$$n_i(T) = 2 \, (k_B T/(2\pi\hbar^2))^{3/2} \, (m_e m_h)^{3/4} \, \exp(-qE_g(T)/(2k_B T))$$

with $\hbar = h/2\pi$, effective electron mass of $m_e$ (assumed to 0.22 $m_0$; $m_0 = 9.10\times10^{-31}$ kg), effective hole mass of $m_h$ (assumed to 0.18 $m_0$), $q = 1.60\times10^{-19}$ C, and bandgap of $E_g(T)$ fitted from $\alpha(\lambda, T)$.

In Fig. 3H, we only show data above 150K for simplicity, considering the phase change and issues the Van Roosbroeck-Shockley Relation has at low temperatures.[8,10]

**Way.2**

The way.1 was successfully applied on MAPbI$_3$,[8] a similar perovskite as FAPbI$_3$. There is a generic method using van-Roosbroeck-Shockley Relation. The core remains that $k_{2r}$ in the Van Roosbroeck–Shockley relation represents the effective thermal equilibrium interaction between the material and its surroundings.

The internal formula is

$$k_{2,r} n_i^2 = \int 4\pi n^2 \alpha \Phi_{BB} dE.$$

However, the escape probability is defined as

$$\eta_{esc} = \frac{\int \pi a \Phi_{BB} dE}{\int 4\pi n^2 \alpha d\Phi_{BB} dE}$$

meaning we can say

$$\eta_{esc} k_{2,r} n_i^2 = \int \pi a \Phi\_BB \, dE$$

where $a$ is the *angle averaged* absorptivity (i.e. the fraction of light absorbed for this specific sample as a function of energy).

$$a(E) = \frac{\int_{\theta=0}^{\frac{\pi}{2}} \int_{\phi=0}^{2\pi} \sin\sin(\theta) \, \cos\cos(\theta) \, a(E,\theta,\phi) d\phi d\theta}{\int_{\theta=0}^{\frac{\pi}{2}} \int_{\phi=0}^{2\pi} \sin\sin(\theta) \, \cos\cos(\theta) \, d\phi d\theta}$$

if we assume that $a(E, \theta, \phi) = a(E, \theta = 0, \phi = 0)$ for all angles then $a(E) = a(E, \theta, \phi)$. One can of course come up with more complex/better models of how absorption changes with angle and then fix these based on what one measures for light perpendicular to the sample.

Either Way.1 or Way.2 gives a $k_{2r}$ value with a multiplier, which is a correction factor due to deviations from $n_i$, as well as the absolute value of absorption coefficient derived from measurements, and uncertainties while dealing with absorption data like the tail below bandgap. One can also introduce more factors, including PL re-absorption and charge-carrier diffusion[8] to reconcile calculated $k_{2r}$ value with measured values. Therefore, it is reasonable to use scaled simulated values via the Van Roosbroeck-Shockley Relation rather than absolute values to avoid introducing more parameters when comparing with the experimental data.

## supplementary Note 3  Fermi-level splitting calculations

The Fermi-level splitting, $\mu_F$ calculation follows our previous work.[11] The key principle of the calculated $\mu_F$ standing for the practical limit of the open circuit voltage is based on the actual available density of excited excess carriers under one sun, determined by the recombination rates in the light-absorber material. The steps to achieve the $\mu_F$ are: firstly, we confirm the first and second order recombination rates, by the global fitting method applied on fluence-dependent TRPL decays recorded by iCCD, as shown in Fig. 1a. The global fitting method from Equation 2 is derived from the normal recombination Equation S1.1. The third order recombination or known as Auger recombination is neglected due to the excitation density lower than $10^{17}$/cm$^3$. The $k_1$ values are checked by the linear fitting on the log-lin plotted trace recorded by using the lowest excitation fluence. Secondly, we replace the generation term in the differential equation of excess carrier density as a function of time from iCCD pulsed laser by the condition of standard continuous one sun illumination at AM1.5. Here, we only consider the number of photons with energies above the bandgap of our material in the AM1.5 spectrum. Thirdly, we extract the density of excess carriers at the plateau of the time-dependent carrier density and input them into the equation of Fermi-level splitting:

$$\mu_F = \frac{KT}{q} ln \frac{(n_0 + \Delta n)(p_0 + \Delta p)}{n_i^2}$$

Here, we assume the materials are p type (equivalent to n type) with background carrier density $p_0$ at room temperature, while at 50K, we treat $p_0$=0 due to the frozen-out effect. The $n_i$ is obtained by equation

$n_i = N_c N_v exp(\frac{-E_g}{kT})$, where $N_c = 2(\frac{2\pi m_n^* kT}{h^2})^{\frac{3}{2}}$ and $m_n^*=m_p^*$=0.4m$_0$ confirmed by ellipsometry[12].

The $E_g$ change on temperature is captured from absorption data.

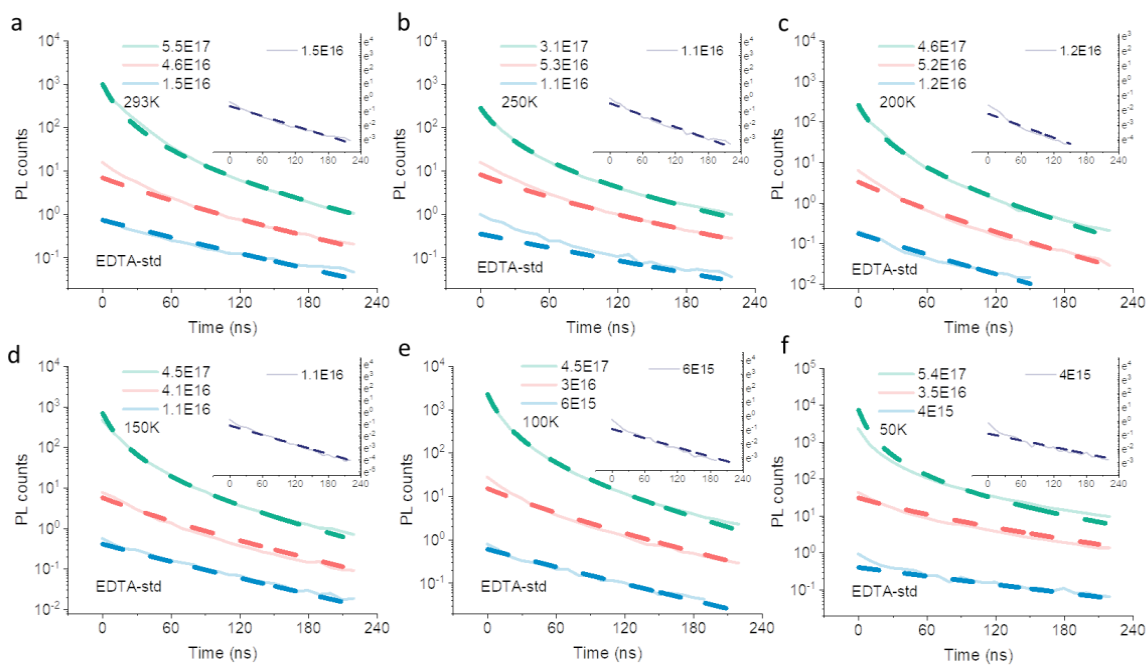

FIG.S3.1 Temperature dependent PL decays and fits of EDTA-std. Insets: linear fits to the decay recorded under the lowest excitation density.

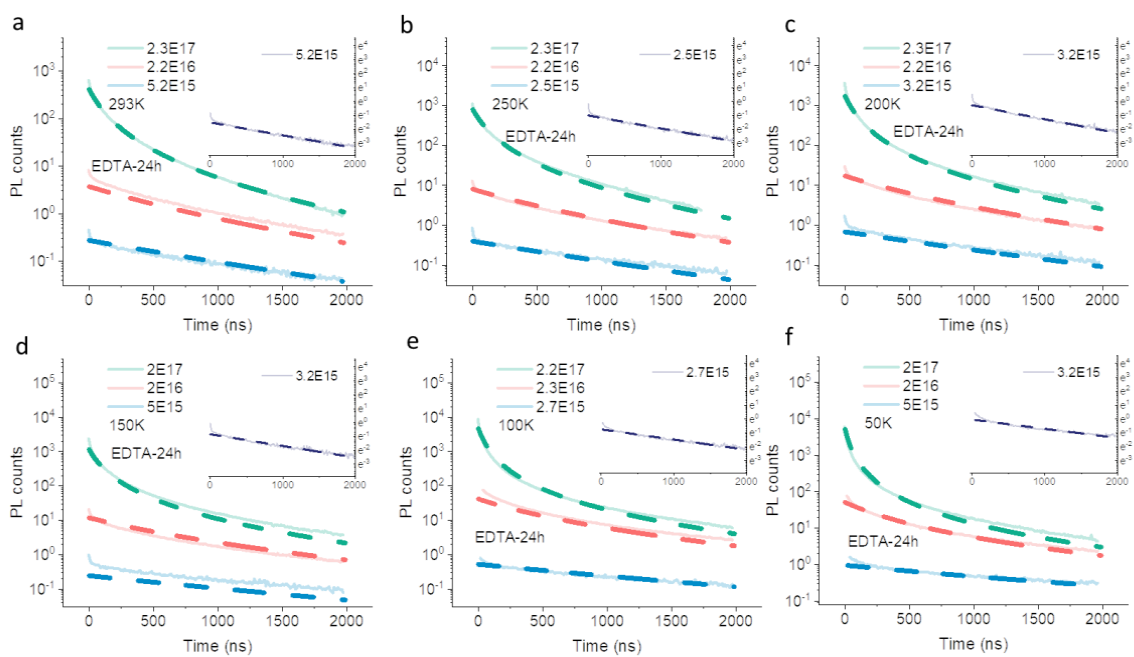

FIG.S3.2 Temperature dependent PL decays and fits of EDTA-24h. Insets: linear fits to the decay recorded under the lowest excitation density.

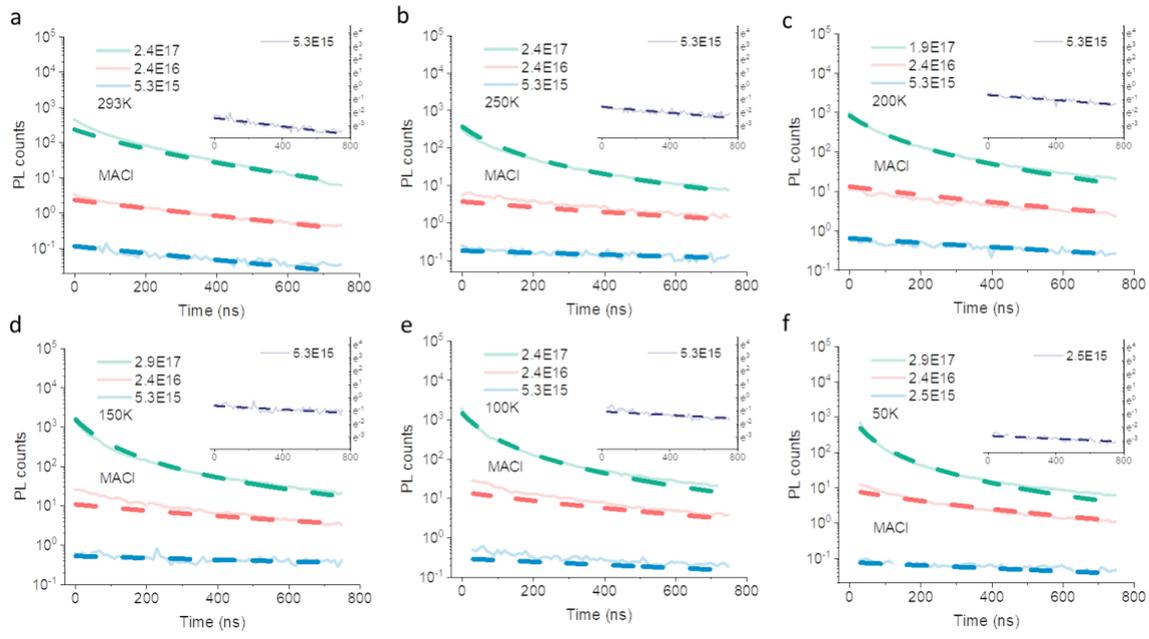

FIG.S3.3 Temperature dependent PL decays and fits of MACl. Insets: linear fits to the decay recorded under the lowest excitation density.

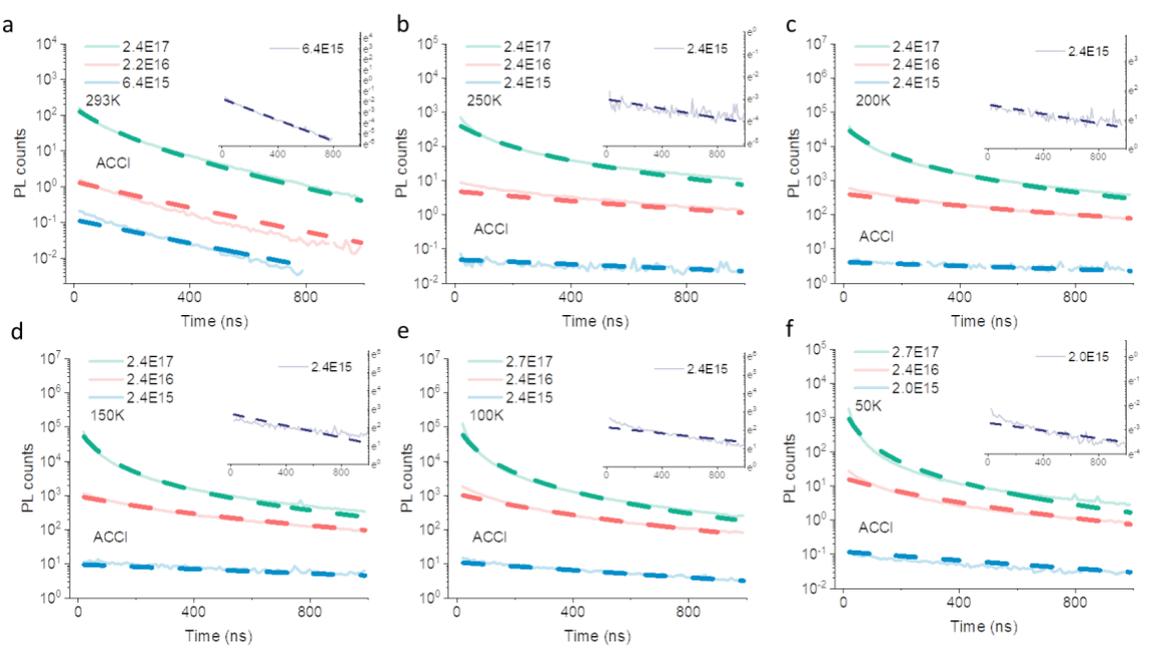

FIG.S3.4 Temperature dependent PL decays and fits of ACCl. Insets: linear fits to the decay recorded under the lowest excitation density.

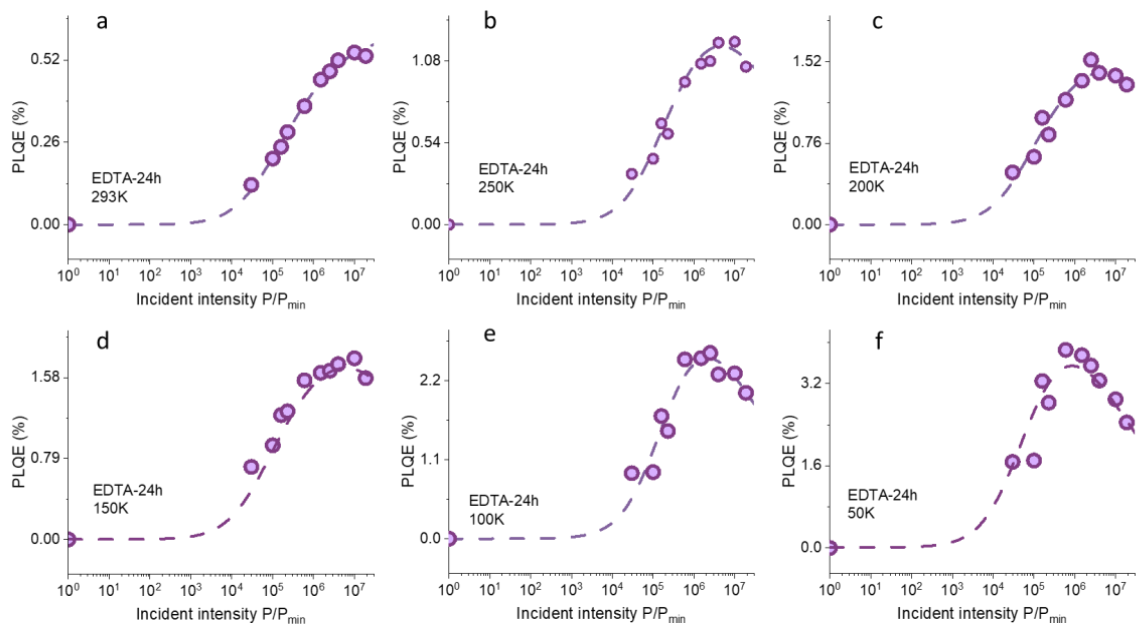

FIG.S4.1 Temperature dependent PLQE and fits of EDTA-24h.

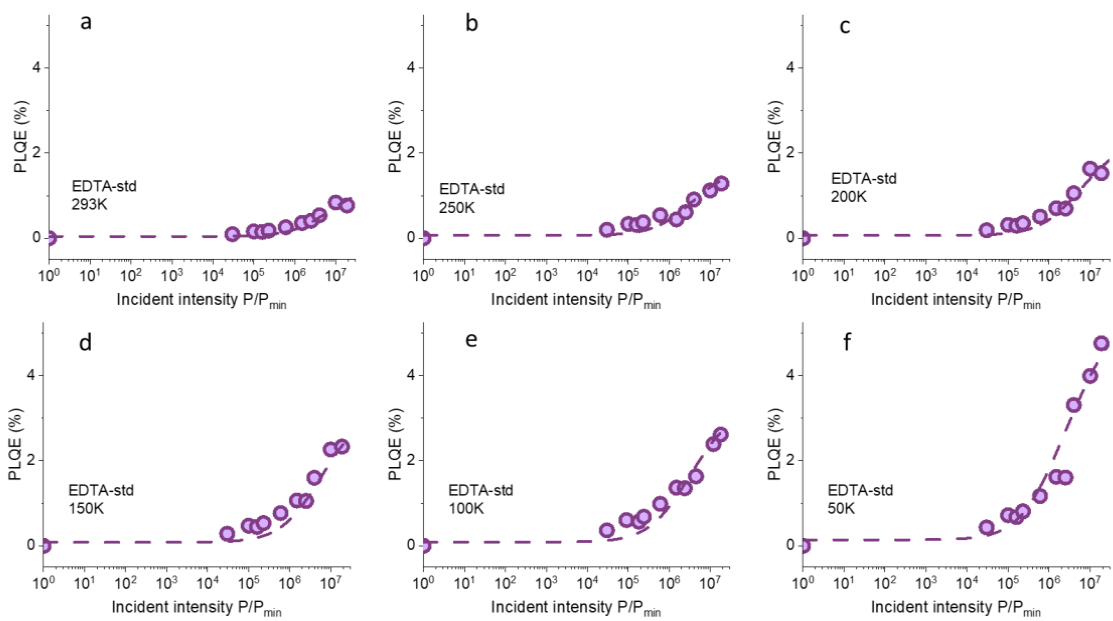

FIG.S4.2 Temperature dependent PLQE and fits of EDTA-std.

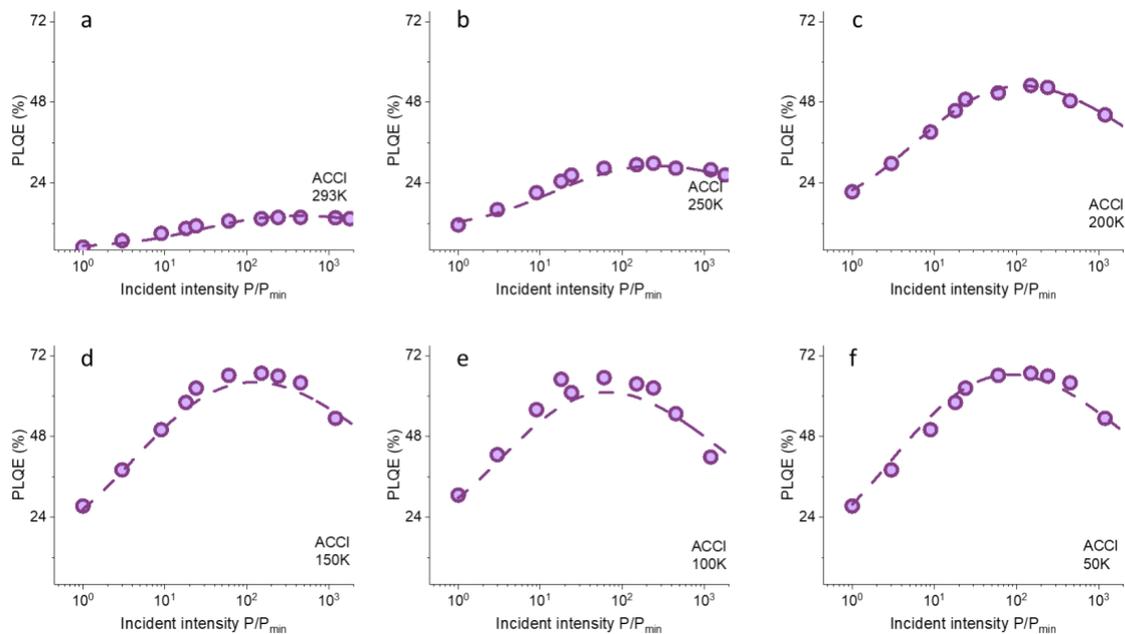

FIG.S4.3 Temperature dependent PLQE and fits of ACCl.

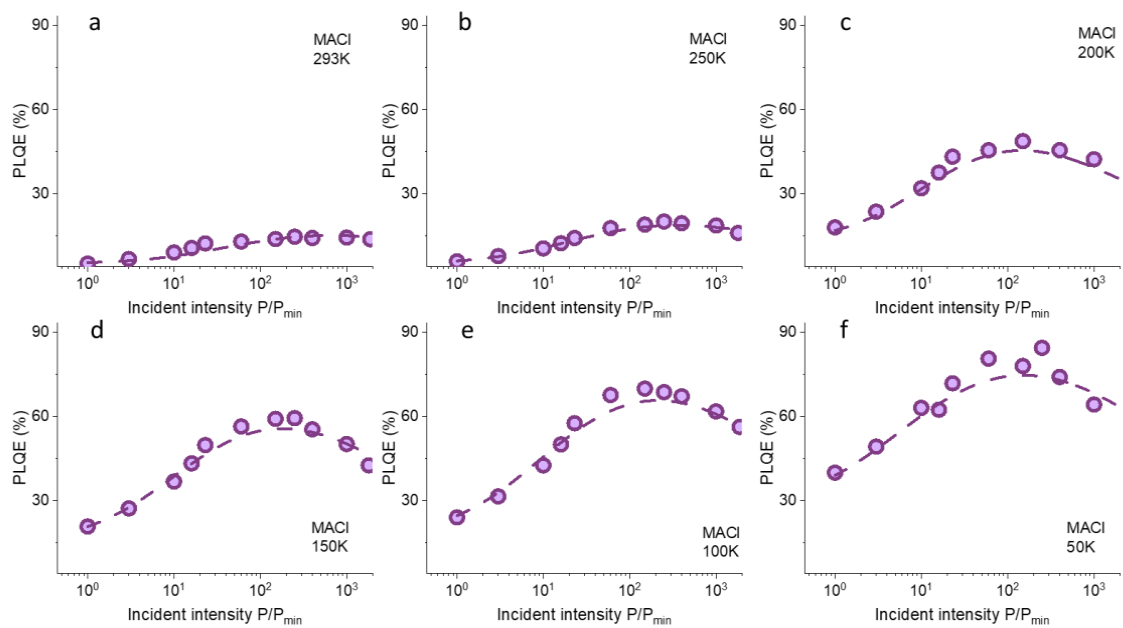

FIG.S4.4 Temperature dependent PLQE and fits of MACl.

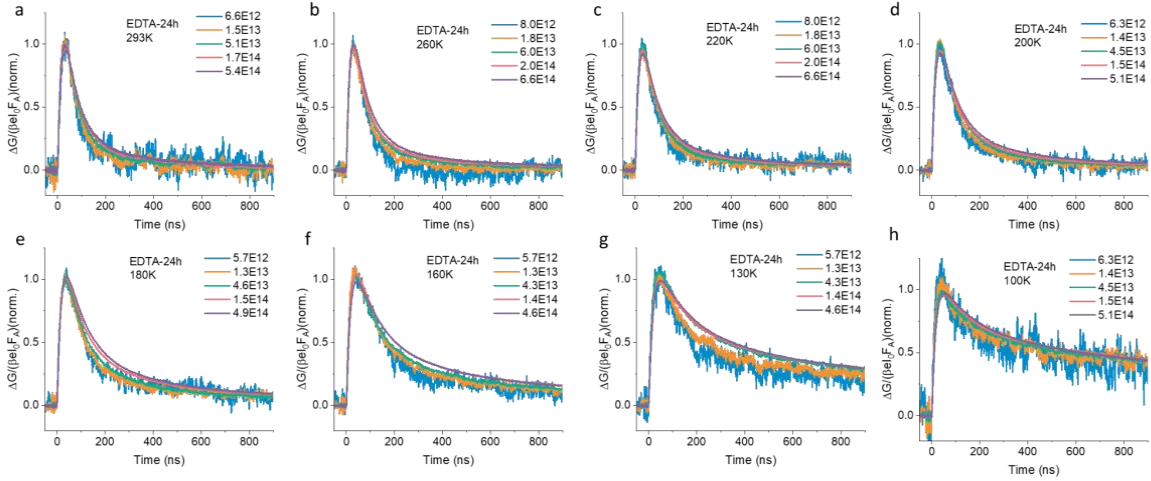

FIG.S5 Temperature dependent TRMC (normalised to one) of EDTA-24h.

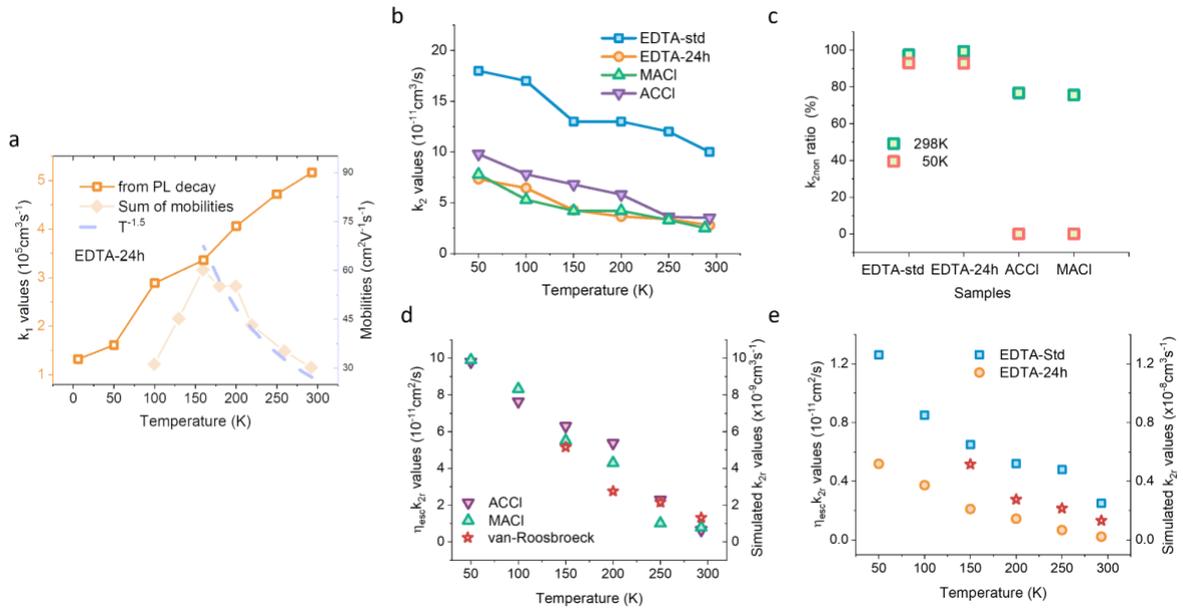

FIG.S6 Temperature dependent $k_1$ values along with the product of generation yield and sum of mobilities. The increase in mobility follows the power law $T^{-1.5}$, showing a standard temperature dependent behaviour of mobilities in a direct bandgap semiconductor, which means the lattice scattering is the major limit reason, supported by the negligible temperature dependence of absorbance spectra in Fig.1D, unlike the $T^{-1.6 \sim -2.3}$ from significant temperature dependence of absorbance spectra of an indirect bandgap semiconductor like Si. (b) direct $k_2$ comparison between different samples from PLQE fits. (c) $k_{2non}$ comparison between different samples from PLQE fits at 298 K and 50K, respectively. The difference is related to precursor doping and surface condition.[1–3] (d,e) The unnormalised comparison between the extracted $\eta_{esc}k_{2r}$ values and simulated values from Van Roosbroeck and Shockley Relation.

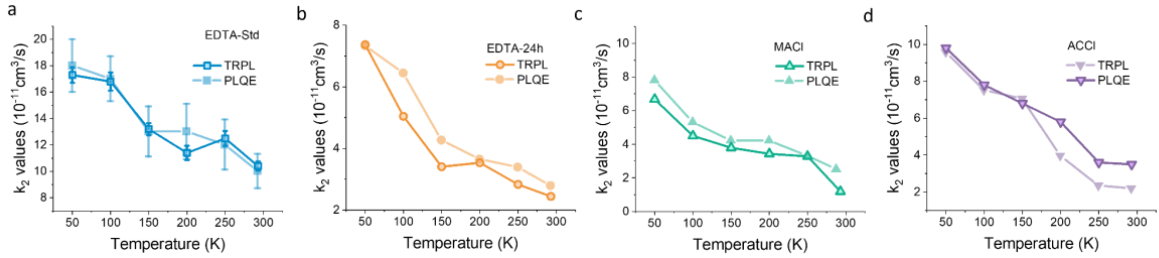

FIG.S7. Trend agreement between the $k_2$ values extracted from TRPL and PLQE at each temperature for each of the four different samples. Error bars are representatively included for the EDTA-std sample, obtained through stochastic fitting, factoring in the variability and uncertainty of the fitting parameters.

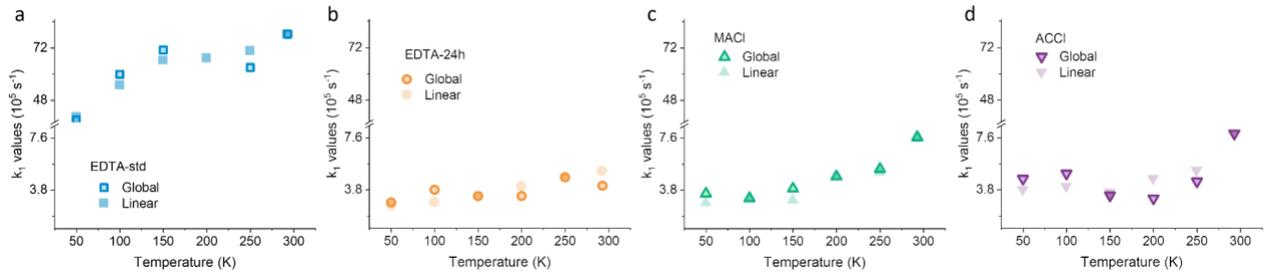

FIG.S8. Temperature dependent $k_1$ values of each sample. The extracted $k_1$ from global fits and linear fits on trace under the lowest excitation are close to each other at each temperature point.

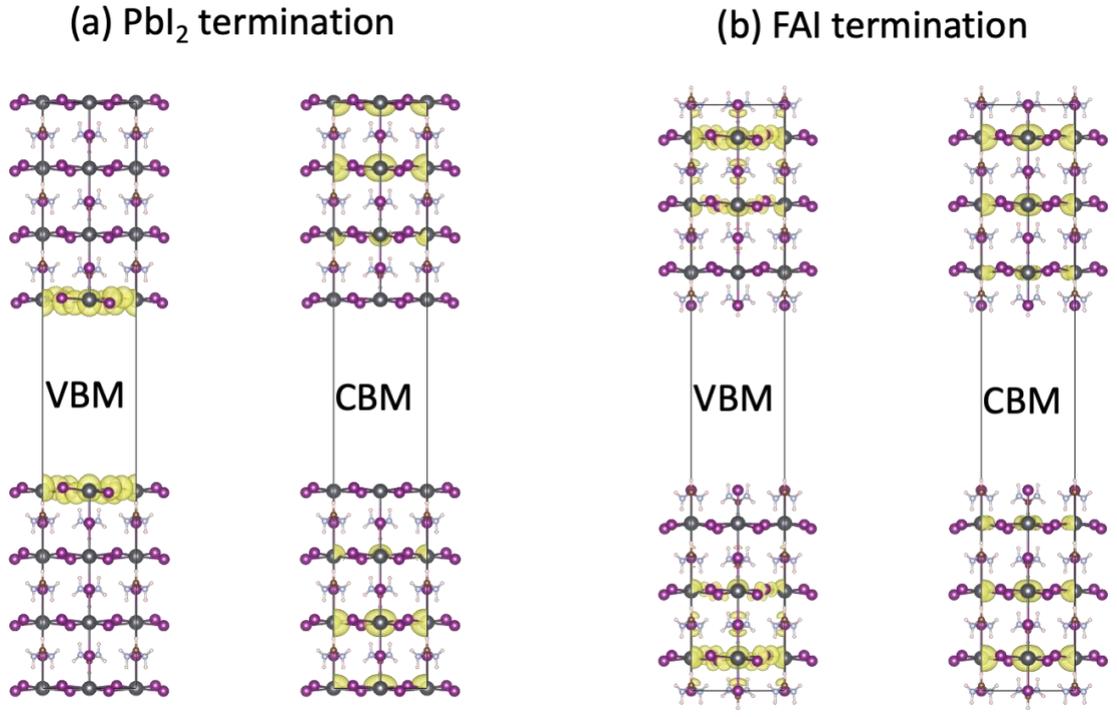

FIG.S9. DFT calculations of ots-FAPbI$_3$ (001) surface. (a) PbI$_2$ termination. [PbI6]4- octahedra are truncated at the surface. Hole carriers will accumulate on the surface and electron carriers will stay in the bulk region. Spontaneous carrier separation will happen. (b) FAI termination. [PbI6]4- octahedra

are not truncated at the surface. Hole carriers will stay in the bulk region and electron carriers will exist across the material. Spontaneous carrier separation will not happen.

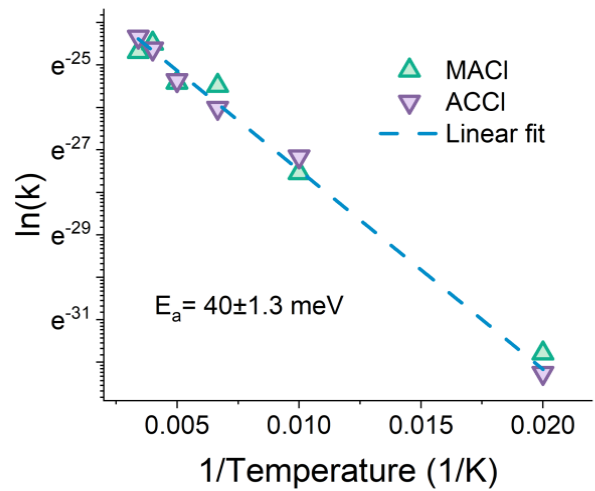

FIG.S10. Arrhenius plot of the natural logarithm of MACl and ACCl versus inverse temperature (1/T). Linear fits (dashed line) based on the Arrhenius equation yield an activation energy ($E_a$) of 40 meV, indicative of a thermally activated carrier recombination mechanism. Arrhenius equation:

$$ln(k) = ln(A) - \frac{Ea}{kB} \cdot \frac{1}{T}$$